\documentclass{elsarticle}
\usepackage{amsmath,a4wide}
\usepackage{amssymb}
\usepackage{graphics}
\usepackage{graphicx}
\usepackage{subfigure}
\usepackage{xspace}
\usepackage{rotating}
\usepackage{verbatim}
\usepackage{longtable}
\usepackage{lscape}
\usepackage{multirow}
\usepackage{afterpage}
\usepackage[breaklinks]{hyperref}
\usepackage{breakurl}
\usepackage[english]{babel}
\usepackage{setspace}
\usepackage{lineno}

\newcommand{\be}{\begin{equation}}
\newcommand{\ee}{\end{equation}}
\newcommand{\ben}{\begin{enumerate}}
\newcommand{\een}{\end{enumerate}}
\newcommand{\bi}{\begin{itemize}}
\newcommand{\ei}{\end{itemize}}
\newcommand{\bbe}{\begin{equation*}}
\newcommand{\eee}{\end{equation*}}
\newcommand{\bber}{\begin{equation*}\textcolor{red}}
\newcommand{\eeer}{\end{equation*}}
\newcommand{\bc}{\begin{center}}
\newcommand{\ec}{\end{center}}
\newcommand{\bea}{\begin{eqnarray}}
\newcommand{\eea}{\end{eqnarray}}
\newcommand{\bem}{\begin{pmatrix}}
\newcommand{\eem}{\end{pmatrix}}
\newcommand{\bbea}{\begin{eqnarray*}}
\newcommand{\eeea}{\end{eqnarray*}}

\newcommand{\bts}{{\emph {BtS}}\xspace}

\newcommand{\mr}[1]{\mathrm{#1}}

\newcommand{\xmax}{{$X_\text{max}$}\xspace}

\def\EeV{\ifmmode {\mathrm{\ Ee\kern -0.1em V}}\else
                   \textrm{Ee\kern -0.1em V}\fi\xspace}%
\def\PeV{\ifmmode {\mathrm{\ Pe\kern -0.1em V}}\else
                   \textrm{Pe\kern -0.1em V}\fi\xspace}%
\def\TeV{\ifmmode {\mathrm{\ Te\kern -0.1em V}}\else
                   \textrm{Te\kern -0.1em V}\fi\xspace}%
\def\MeV{\ifmmode {\mathrm{\ Me\kern -0.1em V}}\else
                   \textrm{Me\kern -0.1em V}\fi\xspace}%
\def\GeV{\ifmmode {\mathrm{\ Ge\kern -0.1em V}}\else
                   \textrm{Ge\kern -0.1em V}\fi\xspace}%
\def\keV{\ifmmode {\mathrm{\ ke\kern -0.1em V}}\else
                   \textrm{ke\kern -0.1em V}\fi\xspace}%
\def\MeV{\ifmmode {\mathrm{\ Me\kern -0.1em V}}\else
                   \textrm{Me\kern -0.1em V}\fi\xspace}%
\def\eV{\ifmmode {\mathrm{\ e\kern -0.1em V}}\else
                   \textrm{e\kern -0.1em V}\fi\xspace}%

\def\offline{\mbox{$\overline{\textrm%
{Off}}$\hspace{.05em}\protect\raisebox{.4ex}%
{$\protect\underline{\textrm{line}}$}}\xspace}

\def \mal      {Malarg\"{u}e\xspace}
\def \pao      {Pierre Auger Observatory\xspace}

\def \gcmsq    {$\text{g~cm}^{-2}$\xspace}

\begin{document}
\setlength{\topmargin}{-12mm}

\begin{frontmatter}

\title{Description of Atmospheric Conditions at the \pao using the \emph{Global Data Assimilation System} (GDAS)}

  \author{
\par\noindent
{\bf The Pierre Auger Collaboration} \\
P.~Abreu$^{75}$, 
M.~Aglietta$^{58}$, 
M.~Ahlers$^{110}$, 
E.J.~Ahn$^{94}$, 
I.F.M.~Albuquerque$^{20}$, 
D.~Allard$^{34}$, 
I.~Allekotte$^{1}$, 
J.~Allen$^{98}$, 
P.~Allison$^{100}$, 
A.~Almela$^{13,\: 9}$
% ...

\par\noindent
$^{1}$ Centro At\'{o}mico Bariloche and Instituto 
Balseiro (CNEA-UNCuyo-CONICET), San Carlos de 
Bariloche, Argentina \\
% ...
$^{9}$ Instituto de Tecnolog\'{\i}as en Detecci\'{o}n y 
Astropart\'{\i}culas (CNEA, CONICET, UNSAM), Buenos Aires,
 Argentina \\
% ...
$^{13}$ Universidad Tecnol\'{o}gica Nacional - Facultad 
Regional Buenos Aires, Buenos Aires, Argentina \\
% ...
$^{20}$ Universidade de S\~{a}o Paulo, Instituto de 
F\'{\i}sica, S\~{a}o Paulo, SP, Brazil \\
% ...
$^{34}$ Laboratoire AstroParticule et Cosmologie 
(APC), Universit\'{e} Paris 7, CNRS-IN2P3, Paris, France 
% ...
$^{58}$ Istituto di Fisica dello Spazio 
Interplanetario (INAF), Universit\`{a} di Torino and 
Sezione INFN, Torino, Italy \\
% ...
$^{75}$ LIP and Instituto Superior T\'{e}cnico, Technical
 University of Lisbon, Portugal \\
% ...
$^{94}$ Fermilab, Batavia, IL, USA \\
% ...
$^{98}$ New York University, New York, NY, USA \\
% ...
$^{100}$ Ohio State University, Columbus, OH, USA \\
% ...
$^{110}$ University of Wisconsin, Madison, WI, USA \\
% ...
 (INST), Hanoi, Vietnam \\
\par\noindent
($\dagger$) Deceased \\
(a) at Konan University, Kobe, Japan \\
(f) now at University of Maryland \\
(h) now at NYU Abu Dhabi \\
(i) now at Universit\'{e} de Lausanne \\
}

  \begin{abstract}
    Atmospheric conditions at the site of a cosmic ray observatory must be known
    for reconstructing observed extensive air showers.  The Global Data
    Assimilation System (GDAS) is a global atmospheric model predicated on
    meteorological measurements and numerical weather predictions.  GDAS
    provides altitude-dependent profiles of the main state variables of the
    atmosphere like temperature, pressure, and humidity. The original data and
    their application to the air shower reconstruction of the Pierre Auger
    Observatory are described. By comparisons with radiosonde and weather
    station measurements obtained on-site in \mal and averaged monthly models,
    the utility of the GDAS data is shown.
  \end{abstract}

\end{frontmatter}


\section{Introduction\label{sec:introduction}}

The \pao~\cite{Abraham:2004dt,Abraham:2009pm} is located near the town of \mal
in the province of Mendoza, Argentina. At the site, at the base of the Andes
mountains, two well-established measurement techniques are combined to measure
extensive air showers with energies above some $10^{17}$~eV. The hybrid detector
consists of a Surface Detector (SD) array and five Fluorescence Detector (FD)
buildings. Each of the slightly more than 1600~SD stations is a water-filled
Cherenkov detector, measuring the secondary particles of air showers that reach
the ground. The detectors of the array are spaced by 1.5~km (750~m in a small
infill area in the western part of the array) and provide the lateral particle
distribution around a shower core. Four FD buildings comprise six telescopes
each and one FD enhancement installation consists of three telescopes. In each
FD telescope, the UV light emitted by excited nitrogen molecules along the
shower track is collected by a large segmented mirror and reflected onto a
camera composed of 440~PMTs. With this measurement, the geometry and the
longitudinal profile of the shower can be obtained.

For the reconstruction of extensive air showers, the optical properties of the
atmosphere at the site of the observatory have to be known. This is particularly
true for reconstructions based on data obtained with the fluorescence
technique~\cite{Abraham:2010}, but also impacts upon data collected with the
surface detectors~\cite{Abraham:2009bc}. The detection of clouds is an important
task of the atmospheric monitoring systems. Clouds can obstruct or --~through
scattering of the intense Cherenkov light~-- amplify the apparent fluorescence
light before it reaches the FD.  To eliminate data recorded in cloudy conditions
from physics analyses, lidar stations and infrared cloud cameras are installed
at each FD station of the Pierre Auger Observatory. These instruments scan the
fields of view of the fluorescence detectors several times per hour during data
taking periods to measure the cloud coverage and the base height of
clouds~\cite{Louedec:2011}.  The vertical profile of the aerosol optical depth
is measured once every hour using vertical laser shots from two facilities near
the center of the array. Using the calibrated laser energy and the amount of
light scattered out of the beam towards the FDs, the amount of aerosols can be
estimated~\cite{Abraham:2010}.  Weather conditions near ground, and the
height-dependent atmospheric profiles of temperature, pressure and water vapor
pressure are relevant for several Auger Observatory measurements. E.g., these
parameters affect the production of fluorescence light by excited nitrogen
molecules at the shower track, and the Rayleigh scattering of the light between
the air shower and detector. Atmospheric conditions are measured by intermittent
meteorological balloon radio soundings. Additionally, ground-based weather
stations measure surface data continuously. The profiles from the weather
balloons were averaged to obtain local models, called (new) \mal Monthly Models
\cite{Abraham:2010}. Since March 2009, the atmospheric monitoring system has
been upgraded with the implementation of a rapid monitoring
system~\cite{Keilhauer:2010}. Part of the new program was the measurement of
atmospheric profiles with radio soundings shortly after the detection of
particularly high-energy air showers, a system called Balloon-the-Shower (\bts).
This enables a high-quality reconstruction of the most interesting events.

However, performing radio soundings and applying these data to air shower
analyses is not straightforward. Very critical aspects are the time of the
weather balloon ascent and the data validity period. Furthermore, performing
radio soundings, in particular within \bts, imposes a large burden on the
collaboration. Therefore, we investigate the possibility of using data from the
Global Data Assimilation System (GDAS), a global atmospheric model, for the site
of the Auger Observatory. The data are publicly available free of charge via
READY (Real-time Environmental Applications and Display sYstem). Each data set
contains all the main state variables with their dependence on altitude with a
validity period of 180~minutes for each data set.

Key aspects of the impact of the profiles of atmospheric state variables on the
development and detection of extensive air showers are discussed briefly
(Sec.~\ref{sec:impact}). We motivate the necessity of more reliable atmospheric
profiles by a discussion about the data validity period of weather balloons
(Sec.~\ref{sec:radiosoundings}), describe the content and processing of the GDAS
data (Sec.~\ref{sec:gdas}) and compare them to local measurements
(Sec.~\ref{sec:gdasVSlocal}). The new atmospheric data are implemented in the
data processing and simulation framework of the Auger Observatory for an
analysis of reconstructed air showers (Sec.~\ref{sec:reco}).

\section{Impact of Atmospheric State Variables on the Development and
Detection of Extensive Air Showers
\label{sec:impact}}

Varying atmospheric conditions in terms of state variables like temperature,
pressure and humidity, may alter the development and, in particular, the
detection of extensive air showers. Here, different aspects relevant to the
analysis of air showers at the Pierre Auger Observatory are discussed.

The air fluorescence emission excited by the passage of an air shower depends on
pressure, temperature, and humidity~\cite{Arqueros:2008}. The collisional
de-excitation of excited nitrogen molecules by other molecules of the atmosphere
like nitrogen, oxygen, and water vapor counteracts the de-excitation of the
molecules via radiation. These quenching processes are pressure and temperature
dependent as described by kinetic gas theory, and dependent on the water vapor
content in air.  Furthermore, the collisional cross sections for
nitrogen-nitrogen and nitrogen-oxygen collisions follows a power law in
temperature, $\sigma \propto T^\alpha$.  Most recent experimental data indicate
a negative exponent $\alpha$. In reconstructions of air shower data from the
Auger Observatory, the fluorescence yield with its dependence on atmospheric
conditions is described using experimental results from the AIRFLY
experiment~\cite{Ave:2008,Bohacova:2009afw}. The absolute calibration of the
main fluorescence emission at 337.1~nm is taken from Nagano et
al.~\cite{Nagano:2004}.  The dependence of the fluorescence yield on atmospheric
conditions translates to an atmospheric dependence of the reconstructed cosmic
ray energy and the depth of shower maximum, the latter being an indicator for
the mass of the primary cosmic ray particle. Even short-term variations of the
atmosphere may introduce noticeable effects on these reconstructed parameters.

Besides the fluorescence emission, the pressure, temperature and humidity
profiles of the atmosphere are important for other aspects of the reconstruction
of data collected by the FD.  These include the conversion between geometrical
altitudes and atmospheric depth; the treatment of Cherenkov emission from air
showers; and the transmission of the produced photons from the air shower to the
FD:

\begin{itemize}

\item The air shower development is governed by the interactions and decays of
the secondary particles. These processes are largely determined by the
atmospheric depth $X$, the total column density of atmospheric matter traversed
by the air shower at a given point. $X$ is calculated by integrating the density
of air from the top of the atmosphere, along the trajectory of the shower
through the gas. The observation of the longitudinal shower profiles by
fluorescence telescopes is based on geometrical altitudes $h$. Thus, geometrical
altitudes must be converted into atmospheric depth by taking into account the
actual air density profile $\rho(h)$ at the site of the Observatory, and the
zenith angle $\theta$ of the trajectory of the shower,
\begin{equation}
  X(h_0) = \frac{1}{\cos\theta}\int^\infty_{h_0}\rho(h) \mr{d}h.
\end{equation}

\item The secondary particles in extensive air showers travel faster than the
speed of light in air. As a result, they induce the emission of Cherenkov light
in a narrow, forward-beamed cone. Some of this light in the UV range may be
--~depending on the shower geometry relative to the FD telescope~-- detected
together with the fluorescence light.  To effectively subtract the Cherenkov
photons from the total number of photons detected, the amount of Cherenkov light
emitted by the air shower must be estimated. The Cherenkov yield depends on the
refractive index $n$ of the air, which itself depends on the wavelength of the
emitted light as well as the temperature, pressure and
humidity~\cite{Owens:1967,Ciddor:2002}. Parameterized formulae for the
refractive index of dry air, CO$_2$ and water vapor are used to calculate a
total refractive index,
\begin{equation}
  n_\mr{tot}-1 = (n_\mr{dry}-1) \cdot \frac{\rho_\mr{dry}}{\rho_\mr{air}}
               + (n_\mr{CO_2}-1) \cdot \frac{\rho_\mr{CO_2}}{\rho_\mr{air}}
               + (n_\mr{w}-1) \cdot \frac{\rho_\mr{w}}{\rho_\mr{air}}.
\end{equation}
The refractive index of each component is weighted with its density, which can
be calculated using the number density and the molar mass of the constituent.
Finally, the effect of the decreasing number density with altitude is
parameterized~\cite{Birch:1993} as a function of pressure $p$ and temperature
$\vartheta$ in $^{\circ}$C,
\begin{equation}
\label{eq:nair}
  n_\mr{air}-1 = (n_\mr{tot}-1) \cdot p \cdot
  \frac{1+p\cdot(61.3-\vartheta)\cdot10^{-10}}{96095.4\cdot(1+0.003661\cdot\vartheta)}.
\end{equation}

\item Between the production of fluorescence and Cherenkov light in the air
shower and the detection at the FD telescope, the light is scattered by
molecules in the atmosphere. The transmission of light depends on the Rayleigh
cross section~\cite{Tomasi:2005},
\begin{equation}
  \sigma_\mr{R}(\lambda,p,T,e) = \frac{24\pi^3}{\lambda^4 \cdot N^2} \cdot
  \left( \frac{n_\mr{air}^2-1}{n_\mr{air}^2+2} \right)^2 \cdot F_\mr{air}(\lambda,p,e).
\end{equation}
where $\lambda$ is the wavelength in m and $N$ the atmospheric molecular
density, measured in molecules per m$^{-3}$. $F_\mr{air}$ is the King factor
that accounts for the anisotropy in the scattering introduced by non-spherical
scatter centers, which depends slightly on pressure and humidity. The refractive
index $n_\mr{air}$ depends on several atmospheric state variables, see
Eq.~\ref{eq:nair}.

\end{itemize}

The last three itemized effects on the reconstruction of extensive air showers
can be taken sufficiently into account by using a proper description of the
atmospheric state, e.g., the local atmospheric monthly models derived from
multi-year meteorological balloon radio soundings (see
Sec.~\ref{sec:radiosoundings}). They affect the reconstruction results of air
shower data, mainly primary energy and position of shower maximum, only by
marginally broadening the uncertainties without any significant systematic
shifts.

However, in the case of the earlier discussed fluorescence emission process and
its atmospheric variability, systematic alterations of the reconstruction
results may be seen together with increased uncertainties, even for short-term
variations of the atmospheric parameters. Finally, after this discussion on
atmospheric influences on FD analysis, it should be noted that uncertainties in
the surface detector signals introduced by varying atmospheric conditions close
to the ground are well understood and quantified~\cite{Abraham:2009bc}.

\section{Validity of Radio Soundings\label{sec:radiosoundings}}

Since August 2002, meteorological radio soundings have been performed above the
Pierre Auger Observatory to measure altitude-dependent profiles of atmospheric
variables, mainly pressure, temperature, and relative humidity. Regular
measurements were done until December 2008 in order to collect data for all
months. After applying selection criteria, 261~profiles from the middle of 2002
until the end of 2008 could be used to build the new Malarg\"ue Monthly
Models~\cite{Abraham:2010}. Starting in March 2009, the radio soundings became
part of the rapid atmospheric monitoring system known as the Balloon-the-Shower
(\bts) program \cite{Keilhauer:2010,Keilhauer:2009icrc1}. A fast online air
shower reconstruction with subsequent quality selections is used to trigger the
launch of a weather balloon by a local technician.

\begin{figure}[tb]
  \centering
  \includegraphics[width=0.7\linewidth]{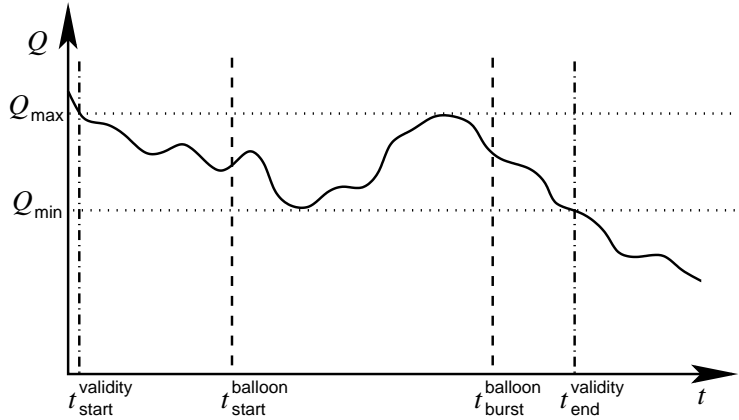}
  \caption{\label{fig:scheme_validity_finding}
    \small{Schematic drawing of the procedure to find an extended time period of
    validity for data from weather balloons based on data measured by
    ground-based weather stations for typical atmospheric conditions. For
    details see text.}
  }
\end{figure}

A procedure was developed to find the period of time for which the data measured
during the ascent of a weather balloon give a good description of the
atmospheric conditions at the Pierre Auger Observatory. The 3-dimensional
atmospheric conditions before and after a weather balloon ascent are unknown but
data from local weather stations may help to identify stable periods or trends
towards rapidly changing conditions. Every active weather station is used as an
independent source of data, no matter how many stations contribute information
during the period of the weather balloon ascent. For each station, the maximum
variations of the temperature, the pressure, and the humidity are obtained for
the duration of the corresponding weather balloon launch defined as the time
between the start of the weather balloon and the burst of the balloon, see
Fig.~\ref{fig:scheme_validity_finding}.

\begin{figure}[htb]
  \begin{minipage} [t]{.33\textwidth}
    \centering
    \includegraphics[width=\linewidth]{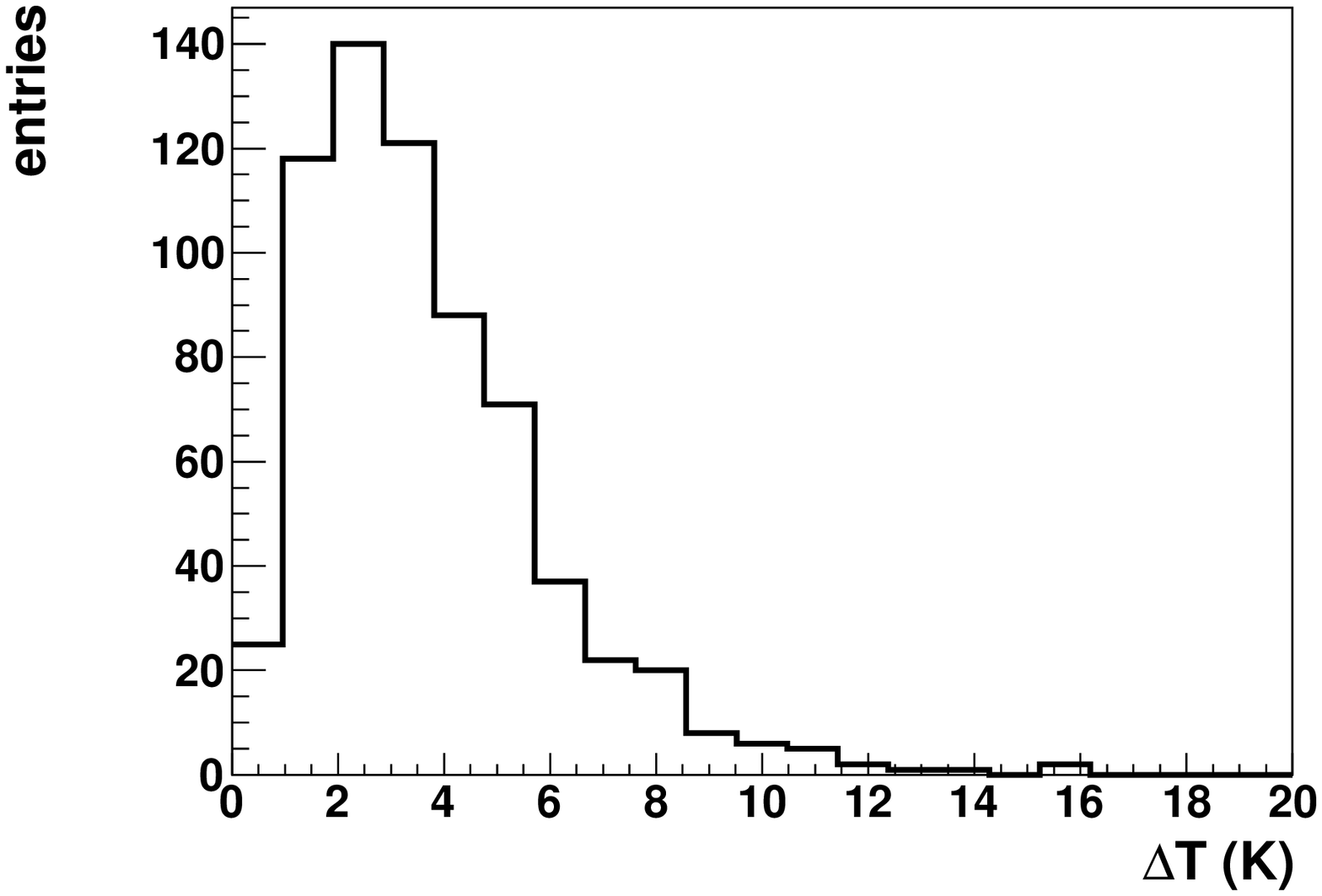}
  \end{minipage}
  \begin{minipage} [t]{.33\textwidth}
    \centering
    \includegraphics[width=\linewidth]{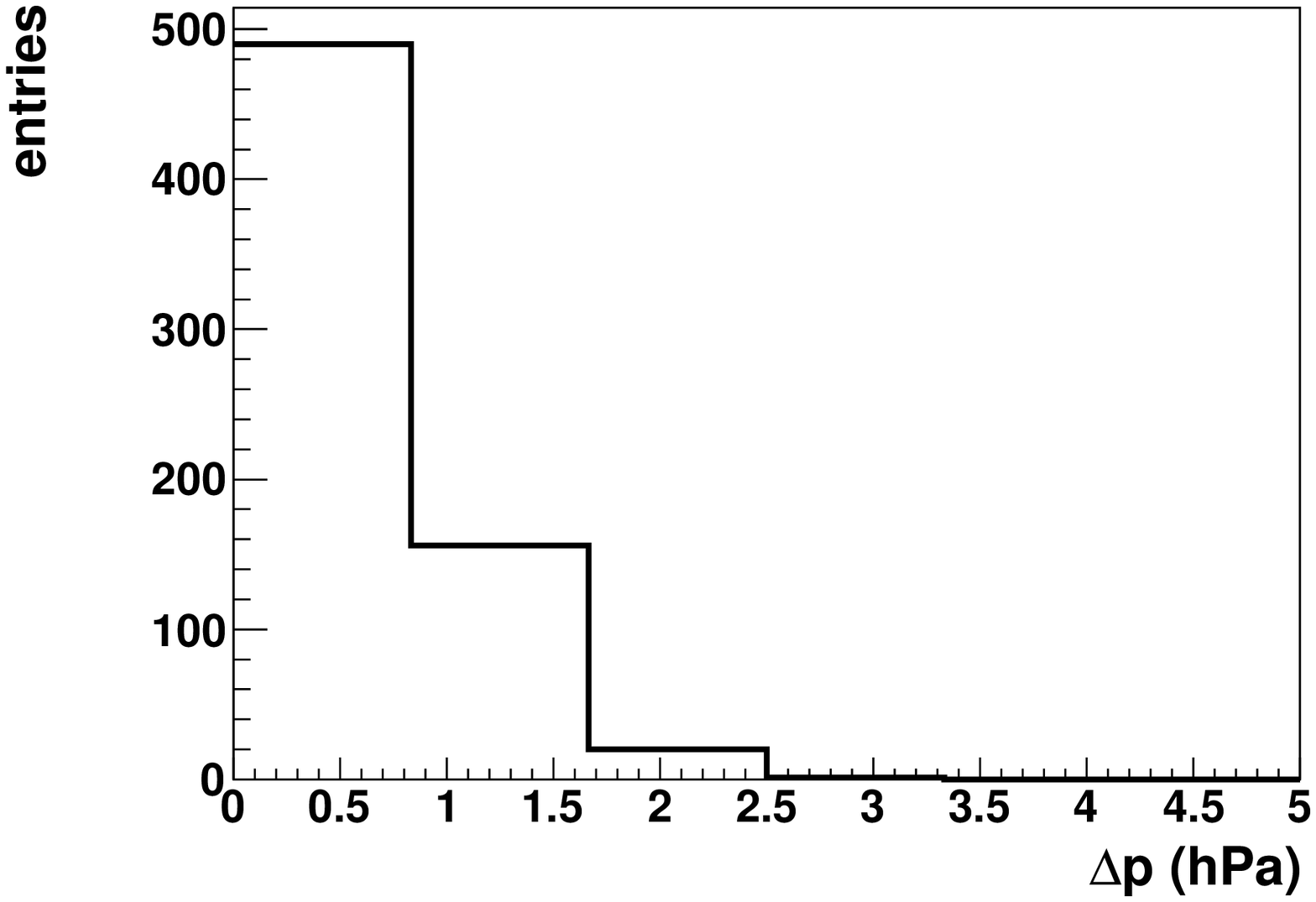}
  \end{minipage}
  \begin{minipage} [t]{.33\textwidth}
    \centering
    \includegraphics[width=\linewidth]{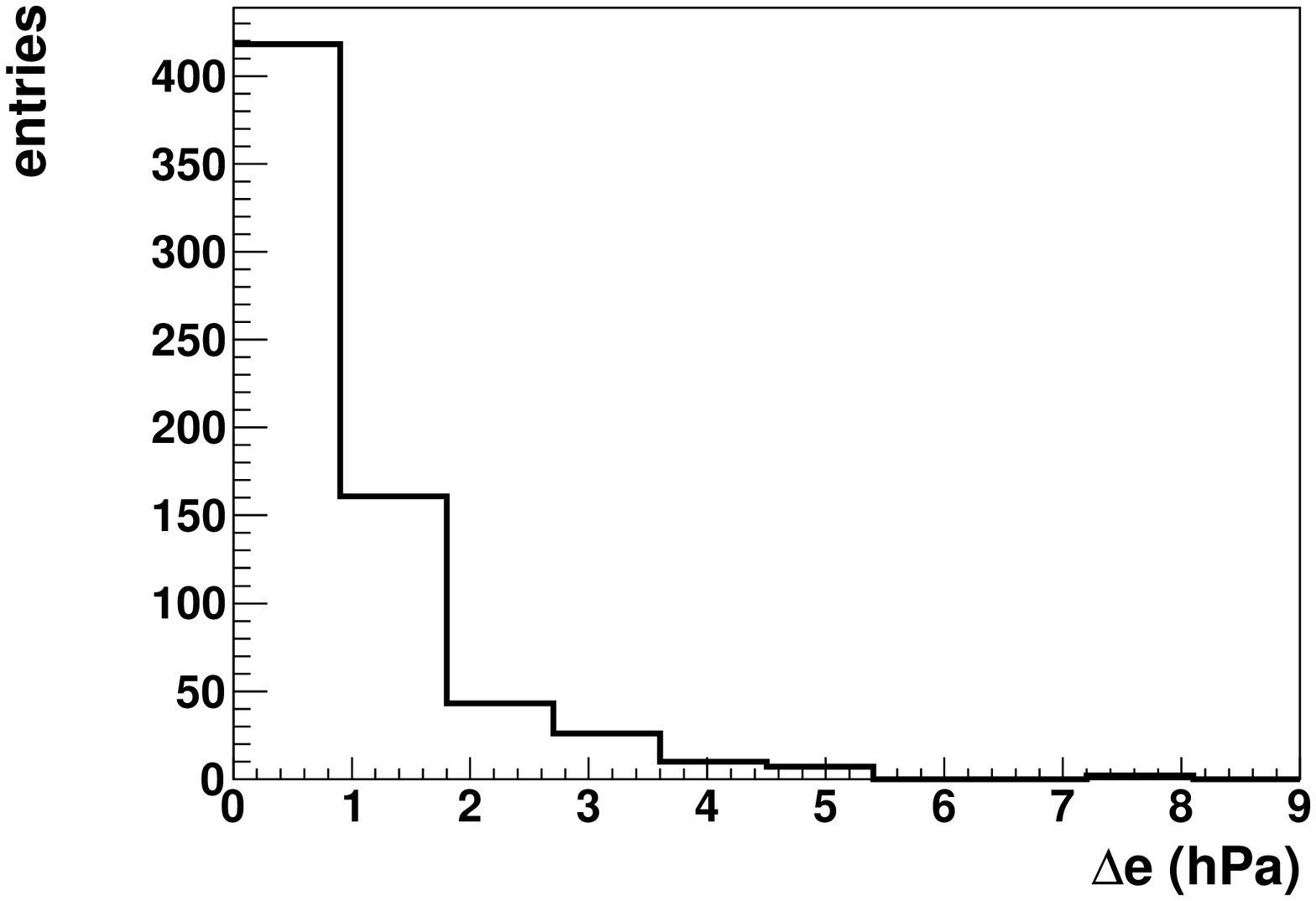}
  \end{minipage}
  \caption{\label{fig:ranges}
    \small{Differences between maximum and minimum values for temperature $T$,
    pressure $p$, and water vapor pressure $e$ as found in weather station data
    during a weather balloon launch. For temperature data (left), typical
    conditions are defined as variances between 4~K and 9~K, with 4~K being
    $\Delta Q_{\rm{low}}$ for the quantity temperature. Periods with $\Delta T <
    4$~K indicate very stable conditions while periods with $\Delta T > 9$~K are
    unstable. For pressure data (middle), typical conditions are defined between
    1 and 3~hPa. For water vapor pressure data (right), the typical conditions
    are between 0.8 and 4.0~hPa. The very stable and unstable conditions of the
    last two quantities are defined accordingly.}
  }
\end{figure}

The difference $\Delta Q$ --~with $Q$ being temperature $T$, pressure $p$, or
water vapor pressure $e$~-- between maximum ($Q_{\rm max}$) and minimum ($Q_{\rm
min}$) values of every station during weather balloon flights can be seen in
Fig.~\ref{fig:ranges}. From these histograms, periods with very stable
conditions ($\Delta Q < \Delta Q_{\rm{low}}$), with typical conditions, and with
unstable conditions are defined for each quantity $Q$ (see caption of
Fig.~\ref{fig:ranges}).

For typical conditions, the data of the weather stations are scanned before and
after the time of the balloon ascent for each quantity $Q$, and the time at which
any quantity leaves the range between $Q_{\rm min}$ and $Q_{\rm max}$ is
determined. This time period gives the validity time period of the radio
sounding for every active weather station. For launches performed during very
stable conditions, the differences in weather station data are quite small.
Thus, only small variations beyond the narrow interval would indicate the end of
validity, imposing very strict cuts on this type of launch. For unstable
conditions, the large $\Delta Q$ values could result in quite long extended
periods of validity. Since both cases result in inappropriate validity periods,
two special criteria for each quantity are found in addition to the typical
case. For very stable conditions, $Q_{\rm max/min}$ are redefined to
$\tilde{Q}_{\rm max/min} = \overline{Q} \pm Q_{\rm low}/2$, where $\overline{Q}$
is the mean of the interval $Q_{\rm min}$ to $Q_{\rm max}$ and $Q_{\rm low}$ is
4~K in the case of temperature data, 1~hPa for pressure data, and 0.8~hPa for
water vapor pressure data.  After definition of $\tilde{Q}_{\rm max/min}$, the
same procedure as for the typical conditions is applied. In case of unstable
conditions, the validity time period is set to the time period during which the
weather balloon ascended.

The average duration of a weather balloon ascent was about 100 minutes. A
validity time period of 200 minutes on average is given by the local weather
station data as described above. Applying this procedure, about half of the
cosmic ray events which triggered the \bts program are observed at times not
covered by the period of validity of the corresponding balloon launch.

Until its termination at the end of 2010, many details of local atmospheric
conditions could be studied with the \bts program. The obtained atmospheric
profiles can be applied to improve the reconstruction of the most interesting,
high-energy air showers. However, the data are not suitable for application to
the standard reconstruction because of their short period of validity. Only very
few air shower events would be covered by atmospheric profiles from radio
soundings.

\section{Global Data Assimilation System (GDAS)\label{sec:gdas}}

In the field of Numerical Weather Prediction, data assimilation is the
adjustment of the development within a model to the real behavior of the
atmosphere as found in meteorological observations~\cite{Mueller:2004}. The
atmospheric models describe the atmospheric state at a given time and position.
Three steps are needed to perform a full data assimilation:

\begin{enumerate}

\item Collect data from meteorological measuring instruments placed all over the
world. These instruments include weather stations on land, ships, and airplanes
as well as radiosondes and weather satellites.

\item Use a short-term forecast from a previous iteration of the numerical
weather prediction together with the measurements to describe the current
situation. This additional information is needed because the available
observations alone are not sufficient. The forecast or \emph{first guess} adds
more information to the system, namely all knowledge of atmospheric behavior
expressed in mathematical model equations.  The models use non-linear
differential equations based on thermodynamics and fluid dynamics.

\item Adjust the model output to the measured atmospheric state. The resulting
3-dimensional image of the atmosphere is called \emph{analysis}.

\end{enumerate}

\begin{figure}[tbp]
  \begin{center}
    \includegraphics[width=.6\linewidth,clip]{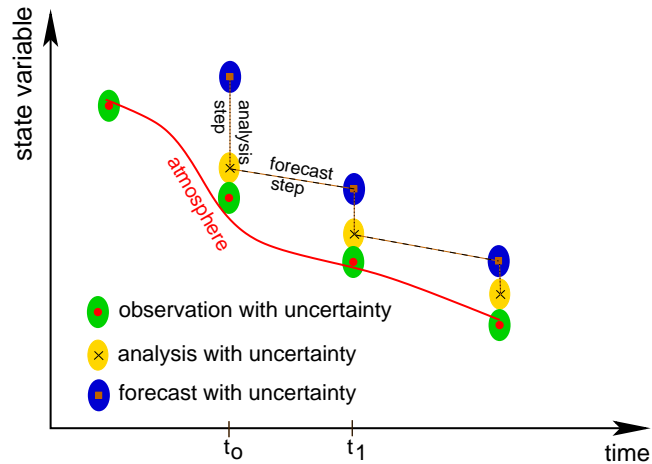}
    \caption{\label{fig:assimilation}
      \small{Schematic of the data assimilation process. Figure adopted in a
      modified form from~\cite{Wergen:2002}.}
    }
  \end{center}
\end{figure}

A schematic showing the principle of data assimilation is given in
Fig.~\ref{fig:assimilation}. At a given time $t_0$, the observations provide the
value of a state variable. A model forecast for this variable from a previous
iteration exists for the same time. The analysis step combines observation and
forecast to describe the current state better than the forecast. This analysis
is the initial point for the weather prediction model to create the forecast for
a later time $t_1$.

\subsection{GDAS Data\label{sec:sourceGDAS}}

The Global Data Assimilation System~\cite{GDASinformation} is an atmospheric
model developed at NOAA's\footnote{National Oceanic and Atmospheric
Administration.} National Centers for Environmental Prediction (NCEP). It
provides an analysis four times a day (0, 6, 12, and 18~UTC) and a 3-, 6- and
9-hour forecast. The numerical weather prediction model used in the GDAS is
the Global Forecast System (GFS).

3-hourly data are available at 23 constant pressure levels -- from 1000~hPa
(roughly sea level) to 20~hPa ($\approx$ 26~km) -- on a global 1$^\circ$-spaced
latitude-longitude grid (180$^\circ$ by 360$^\circ$). Each data set is
complemented by data for the surface level. The data are stored in weekly files
and made available online~\cite{GDASinformation}. In
Table~\ref{t:GDAS_verticallevels}, the level indices corresponding to each data
level are listed. For reference, the altitude from the US Standard Atmosphere
1976 (USStdA)~\cite{USStdA:1976} is also given in the table. The actual height
of the pressure level is stored in the data file. GDAS data are available
starting January 2005. There are two periods without data in the sets.  The
first two weeks of May 2005 and weeks 3 and 4 of November 2005 are missing.
Other than that, the record is complete up to the present time (end of November
2011).

\begin{figure}[tbp]
  \begin{center}
    \includegraphics[width=.74\linewidth,clip]{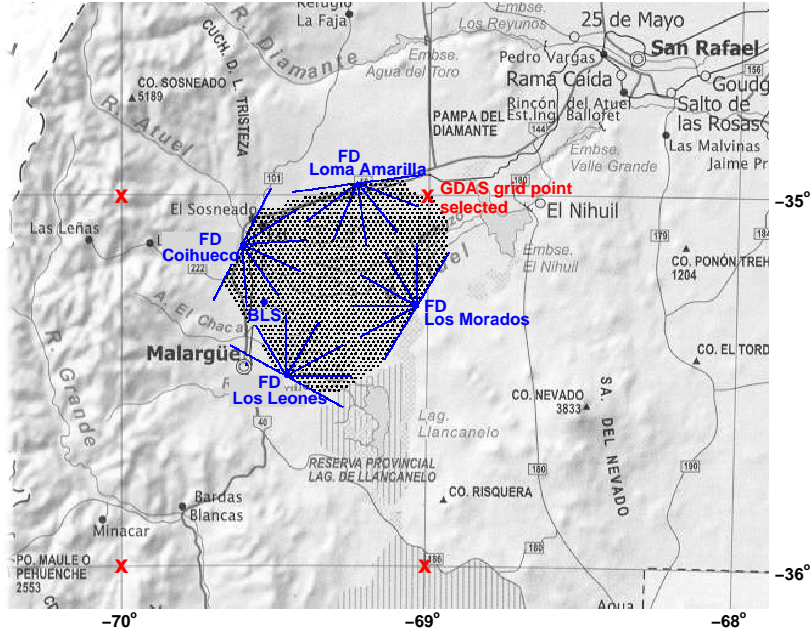}
    \caption{\label{fig:gridPoints}
      \small{Map of the region around the array of the Pierre Auger Observatory
      with geographical latitudes and longitudes marked at the boundary of the
      figure. The positions of the surface and fluorescence detectors are
      superimposed. The available GDAS grid points for that area are marked as
      red crosses. The coordinates of \mal are $-35.48^{\circ}$ (south) and
      $-69.58^{\circ}$ (west), the selected GDAS grid point is $-35^{\circ}$
      (south) and $-69^{\circ}$ (west).}
    }
  \end{center}
\end{figure}

Because of the lateral homogeneity of the atmospheric variables across the Auger
array~\cite{Abraham:2010}, only one location point is needed to describe the
atmospheric conditions. In Fig.~\ref{fig:gridPoints}, the available GDAS grid
points are marked as red crosses on a map together with a map of the surface and
fluorescence detectors of the Auger Observatory. The grid point at 35$^{\circ}$
south and 69$^{\circ}$ west was chosen, at the north-eastern edge of the surface
detector array. The two points to the west of the array are in the foothills of
the Andes mountains and therefore not suitable. The point to the south-east of
the array is quite far away and with a surface height of 1685~m a.s.l., it is
also too high. Nevertheless, the profiles at this point are very similar to
those at the chosen point, on average differing by less than 1~$^\circ$C in
temperature and less than 0.3~hPa in water vapor pressure at all altitudes,
confirming the homogeneity.

\begin{table}[tbp]
  \begin{center}
    \caption{\label{t:GDAS_verticallevels}
      \small{Vertical levels of pressure surfaces. The actual height for each
      pressure level is given in the data file. For reference, the USStdA height
      is also listed in this table. Pressure level~0 contains the surface
      values.}
    }
    \vspace{10pt}
    \begin{tabular}{l l l | l l l l}
      \hline
      Level & Pressure [hPa] & Height [km] & &
      Level & Pressure [hPa] & Height [km] \\
      \hline
      23    & 20   & 26.4 & &    11    & 600  &  4.2 \\
      22    & 50   & 20.6 & &    10    & 650  &  3.6 \\
      21    & 100  & 16.2 & &    9     & 700  &  3.0 \\
      20    & 150  & 13.6 & &    8     & 750  &  2.5 \\
      19    & 200  & 11.8 & &    7     & 800  &  1.9 \\
      18    & 250  & 10.4 & &    6     & 850  &  1.5 \\
      17    & 300  &  9.2 & &    5     & 900  &  1.0 \\
      16    & 350  &  8.2 & &    4     & 925  &  0.8 \\
      15    & 400  &  7.2 & &    3     & 950  &  0.5 \\
      14    & 450  &  6.3 & &    2     & 975  &  0.3 \\
      13    & 500  &  5.6 & &    1     & 1000 &  0.1 \\
      12    & 550  &  4.9 & &    0     & surface &   \\
      \hline
    \end{tabular}
  \end{center}
\end{table}

\begin{figure}[htbp]
  \begin{center}
    \includegraphics*[width=.49\linewidth,clip]{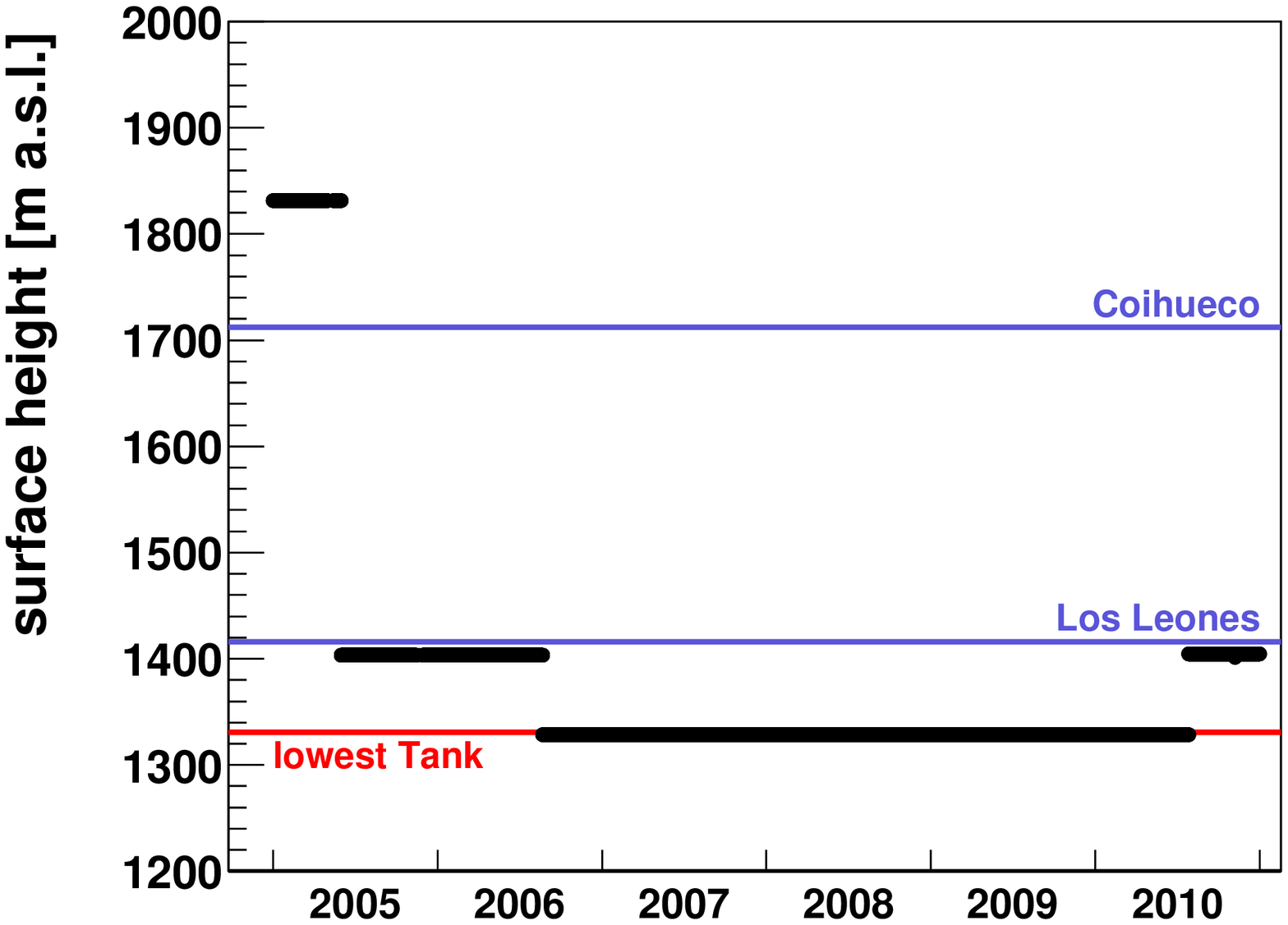}
    \caption{\label{fig:surfaceHeight}
      \small{The surface height of the GDAS data between January 2005 and December
      2010. For reference, the red line marks the height of the lowest tank in
      the array, and the blue lines represent the height of the FD buildings at Los
      Leones and Coihueco.}
    }
  \end{center}
\end{figure}

The height at which the surface data are given changes over the years for the
selected grid point. Starting in January 2005, the surface altitude is 1831.29~m
above sea level. On May~31, 2005, the surface height changes to 1403.38~m, and
on August~22, 2006 it goes down further to 1328.68~m and stays within a few
centimeters of this value until July~27, 2010, when it changes to 1404.65~m.  In
Fig.~\ref{fig:surfaceHeight}, the surface height provided by the GDAS data sets
is shown between January 2005 and December 2010. For reference, the altitudes of
the lowest SD tank (1331.05~m) and the highest and lowest FD buildings (1712.3~m
and 1416.2~m) are also shown. The reasons for these changes are regular
improvements of the models and calculations used to produce the GDAS profiles,
or resolution changes in the meteorological model~\cite{GDASchanges}.  These
changes can occur again in the future, so the surface height of the data has to
be monitored for undesired changes~\cite{StunderPriv:2011}.

For air shower analysis, only data above ground level in \mal are interesting.
Therefore, we only use data from the surface and from pressure levels~6 and
above. The data from beginning of January to the end of May 2005 have a surface
height of around 1800~m. This is even above the height of the highest FD
building at Coihueco. We decided not to attempt an extrapolation down to the
actual ground level of around 1300~m and discard these data. Therefore, the
first data set we use is from June~1, 2005 at 0:00~UTC.

\subsection{Preprocessing of Data\label{sec:preproc}}

For air shower analyses, several types of information are stored in databases
such as the one describing the state variables of the atmosphere. It contains
values for temperature, pressure, relative humidity, air density, and
atmospheric depth at several altitude levels. The first three quantities and the
altitude are directly available in the GDAS data. Air density and atmospheric
depth must be calculated. The surface data contain ground height, pressure at
the ground, and relative humidity 2~meters above ground. Two temperature values
are given, one at the surface and one 2~meters above ground. We decided to use
the latter since we use it together with the relative humidity, which is also
given 2~meters above ground, to calculate water vapor pressure.

In the GDAS data, the altitude is given in geopotential meters with respect to a
geoid (mean sea level). In the air shower analysis framework of the Auger
Observatory, geometric heights with respect to the WGS-84 ellipsoid are used. To
move from geoid to ellipsoid, a constant value of 26~m that arises from the
geographic location of the Auger Observatory must be added to the height values
of the model. The second step is to convert from geopotential height~$h$ to
geometric altitude~$z$ (both measured in m),
\begin{equation}
  z(h,\Phi) = (1+0.002644\cdot\cos(2\Phi))\cdot h
            + (1+0.0089\cdot\cos(2\Phi))\cdot\frac{h^2}{6245000},
\end{equation}
where $\Phi$ is the geometric latitude~\cite{geopot}.

To calculate the air density, the relative humidity must be converted into water
vapor pressure, the partial pressure of water in air in Pa. This conversion
depends on air temperature. The following approximation of the empirical Magnus
formula is used in these calculations:
\begin{eqnarray}
  e = \frac{u}{100\%} \cdot 6.1070 \cdot \mr{exp}\left(
      \frac{17.15 \cdot \vartheta}{234.9 + \vartheta}
      \right),~~~\vartheta~\ge~0^{\circ}\mr{C}, \\
  e = \frac{u}{100\%} \cdot 6.1064 \cdot \mr{exp}\left(
      \frac{21.88 \cdot \vartheta}{265.5 + \vartheta}
      \right),~~~\vartheta~<~0^{\circ}\mr{C},
\end{eqnarray}
where $u$ is the relative humidity in \% and $\vartheta$ is the temperature in
$^{\circ}$C. Now, the air density in kg m$^{-3}$ can be calculated with
\begin{equation}
  \rho = \frac{p \cdot M_\mr{air}}{R \cdot T},
\end{equation}
where $p$ is the pressure in Pa, $T$ is the temperature in Kelvin, $M_\mr{air}$
is the molar mass of air in kg mol$^{-1}$ and $R$ is the universal gas constant
8.31451~J~K$^{-1}$~mol$^{-1}$. Moist air can be separated into three components
to calculate its molar mass: dry air, water vapor and carbon dioxide. The molar
mass of moist air is the sum of the molar masses of the components, weighted
with the volume percentage of that component,
\begin{equation}
  M_\mr{air} = M_\mr{dry} \cdot \varphi_\mr{dry}
             + M_\mr{w} \cdot \varphi_\mr{w}
             + M_\mr{CO_2} \cdot \varphi_\mr{CO_2}.
\end{equation}

The molar masses for dry air, water vapor and CO$_2$ are 0.02897, 0.04401 and
0.01802 kg~mol$^{-1}$, respectively. The volume percentage of CO$_2$ is taken as
385~ppmv, the percentage of water $\varphi_\mr{w}$ is the partial pressure $e$
of water vapor divided by the pressure $p$, and dry air makes up the rest,
$\varphi_\mr{dry} = 1 - \varphi_\mr{w} - \varphi_\mr{CO_2}$. The atmospheric
depth can be calculated by integrating the air density in kg m$^{-3}$ downward
along a vertical track in the atmosphere, starting at infinity
\begin{equation}
\label{eq:depthInt}
  X(h_0) = \int_{h_0}^\infty \rho~\mr{d}h.
\end{equation}
Here, the atmospheric depth is calculated in kg m$^{-2}$. In the astroparticle
community, the unit g cm$^{-2}$ is more common and will be used in this paper.

Since the GDAS data only go up to around 25 to 30~km, we have to approximate the
atmospheric depth at the top of the data profile using $X = p/g$ and integrate
numerically from that height down to ground level. $g$ is the gravitational
acceleration with dependence on altitude and geographical
latitude~\cite{Bodhaine:1999},
\begin{eqnarray}
\nonumber
\label{eq:gravAcc}
  g(h,\Phi) = g_0(\Phi) &-& (3.085462 \cdot 10^{-4} + 2.27 \cdot 10^{-7} \cdot \cos(2\Phi)) \cdot h\\ \nonumber
                        &+& (7.254 \cdot 10^{-11} + 1.0 \cdot 10^{-13} \cdot \cos(2\Phi)) \cdot h^{2}\\
                        &-& (1.517 \cdot 10^{-17} + 6 \cdot 10^{-20} \cdot \cos(2\Phi)) \cdot h^{3},
\end{eqnarray}
where $g$ is in cm~s$^{-2}$ and $g_0(\Phi)$ is the acceleration at sea level
with dependence on latitude,
\begin{equation}
\label{eq:gravAccSeaLvl}
  g_0(\Phi) = 980.6160 \cdot (1 - 0.0026373 \cdot \cos(2\Phi) + 0.0000059 \cdot \cos^{2}(2\Phi)).
\end{equation}
The integration is done by interpolating the density every 200~meters and using
the trapezoidal rule to approximate the integral.

For the simulation and reconstruction of air showers, the description of the
atmospheric parameters should ideally range from ground level to the top of the
atmosphere. GDAS provides data between about 1400 and 30\,000~m. All profiles
are extended up to 100\,000~m --~the approximate boundary to outer space~--
using the US Standard Atmosphere 1976~\cite{USStdA:1976} which describes the
conditions above 30\,000~m reasonably well. Below 1400~m, pressure, atmospheric
depth, density, and water vapor pressure are exponentially extrapolated down to
1000~m based on the lowest two data points. The temperature profile is
extrapolated linearly.  Both extensions are outside the field of view of all FD
stations.

\section{GDAS vs.\ Local Measurements\label{sec:gdasVSlocal}}

To validate the quality of the GDAS data and to verify its applicability for air
shower reconstructions at the Auger Observatory, GDAS data are compared with
local measurements -- atmospheric soundings with weather balloons and
ground-based weather stations. The new \mal Monthly Models (nMMM) are also shown
in some comparisons as a reference since they were the standard profiles used in
reconstructions until recently.

\subsection{GDAS vs.\ Soundings with Weather Balloons
\label{sec:GDASvsRadio}}

Local atmospheric soundings have been performed above the array of the
observatory since 2002, but not on a regular basis. In the beginning, several
week-long campaigns of launches were performed. Then, a pre-determined schedule
was used to coordinate launches during and between dark periods with FD data
taking and finally, with the \bts program, soundings are triggered by
particularly high-energy events during FD data taking only. Most of the balloons
were launched from the Balloon Launching Station (BLS, see
Fig.~\ref{fig:gridPoints}). To provide a set of atmospheric data for every
measured air shower, the soundings were averaged to form monthly mean profiles.
The current version of these \mal Monthly Models (nMMM) were compiled in early
2009. The models on average describe the atmosphere reasonably well, but show
considerable fluctuations when compared to the actual sounding
data~\cite{Abraham:2010}. The uncertainties of the profile for each variable are
given by the standard error of the variation within each month together with the
absolute uncertainties of the sensors measuring the corresponding quantity. For
atmospheric depth profiles, a piecewise fitting procedure is performed to ensure
a reliable application of these parameterizations to air shower simulation
programs. An additional uncertainty is included which covers the quality of the
fitting procedure.

\begin{figure}[htbp]
  \begin{minipage}[t]{.49\textwidth}
    \centering
    \includegraphics*[width=.99\linewidth,clip]{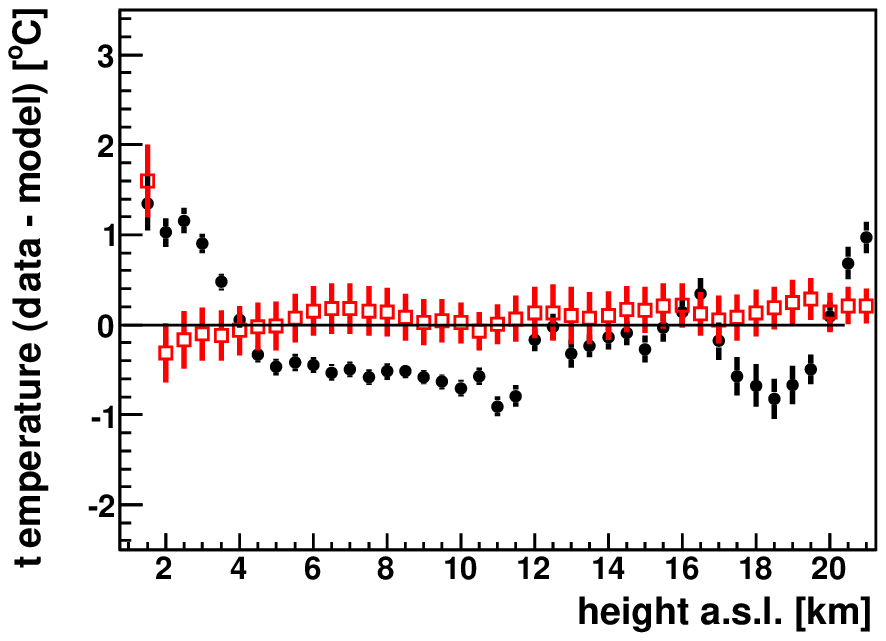}
  \end{minipage}
  \hfill
  \begin{minipage}[t]{.49\textwidth}
    \centering
    \includegraphics*[width=.99\linewidth,clip]{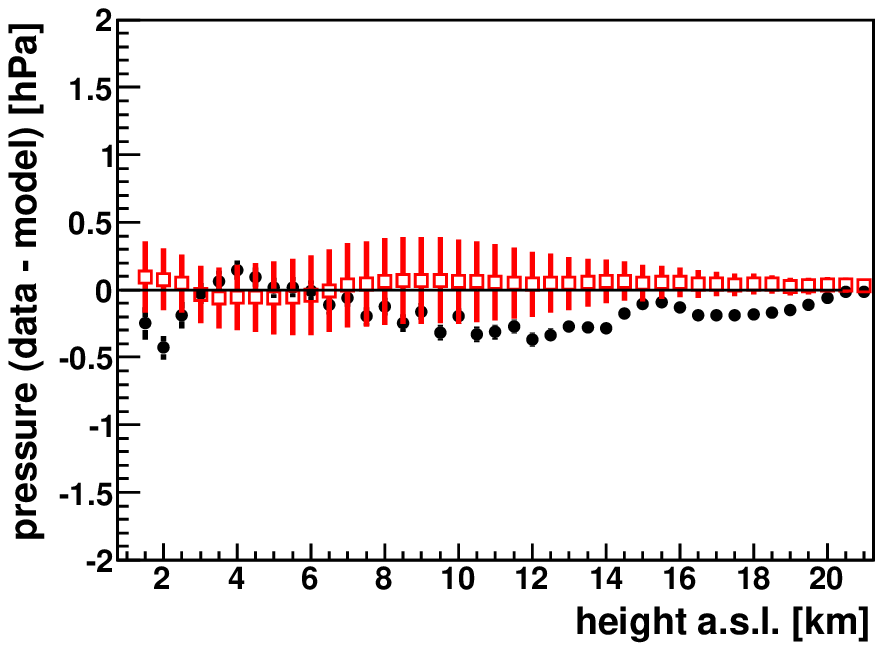}
  \end{minipage}
  \begin{minipage}[t]{.49\textwidth}
    \centering
    \includegraphics*[width=.99\linewidth,clip]{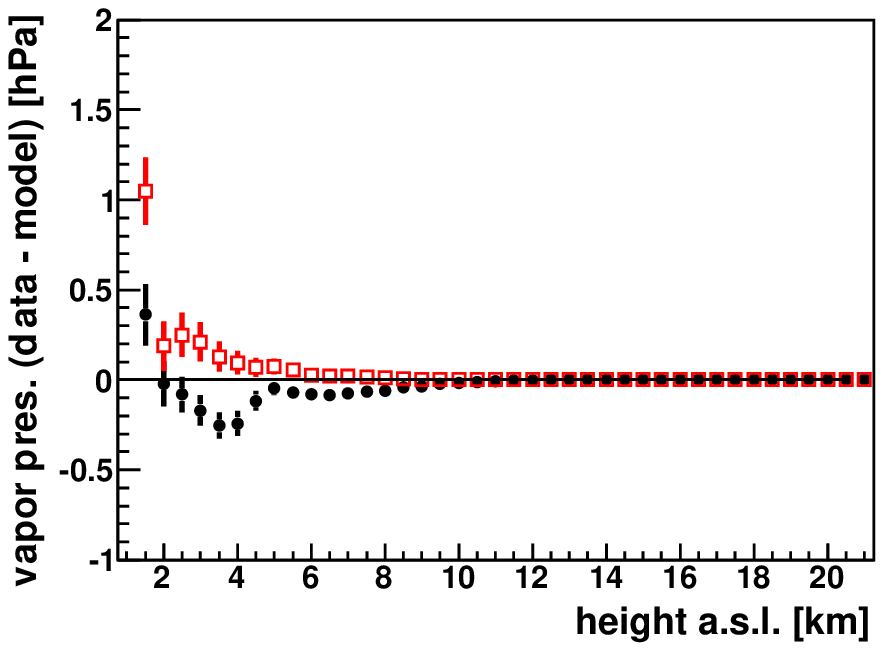}
  \end{minipage}
  \hfill
  \begin{minipage}[t]{.49\textwidth}
    \centering
    \includegraphics*[width=.99\linewidth,clip]{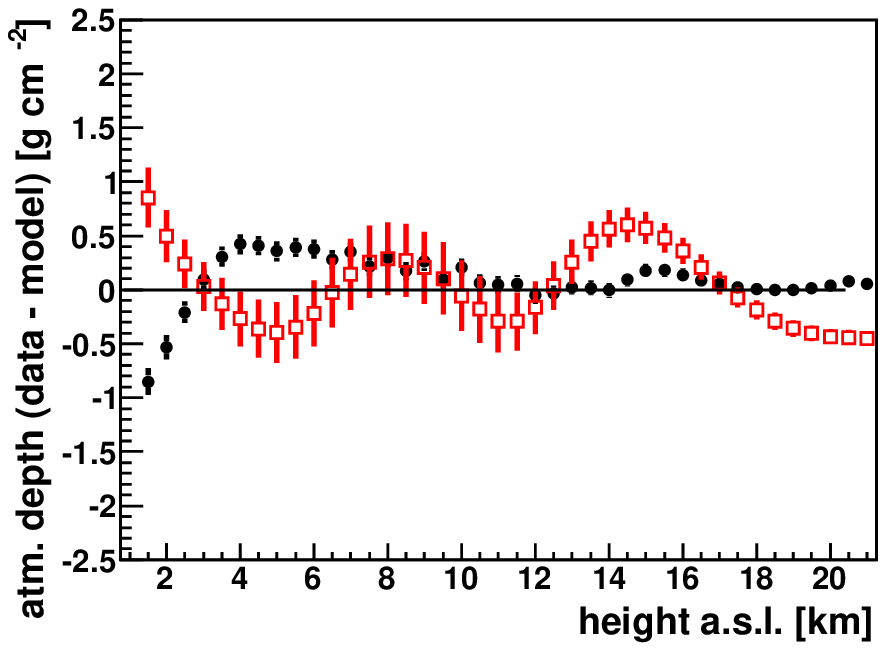}
  \end{minipage}
 \vspace{-12pt}
  \caption{\label{fig:SondeVsGDAS_nMMM}
    Difference of measured radiosonde data and GDAS models (black dots) and
    monthly mean profiles (red squares) versus height for all sounding data
    between 2005 and 2008.
  }
 \vspace{12pt}
  \begin{minipage}[t]{.49\textwidth}
    \centering
    \includegraphics*[width=.99\linewidth,clip]{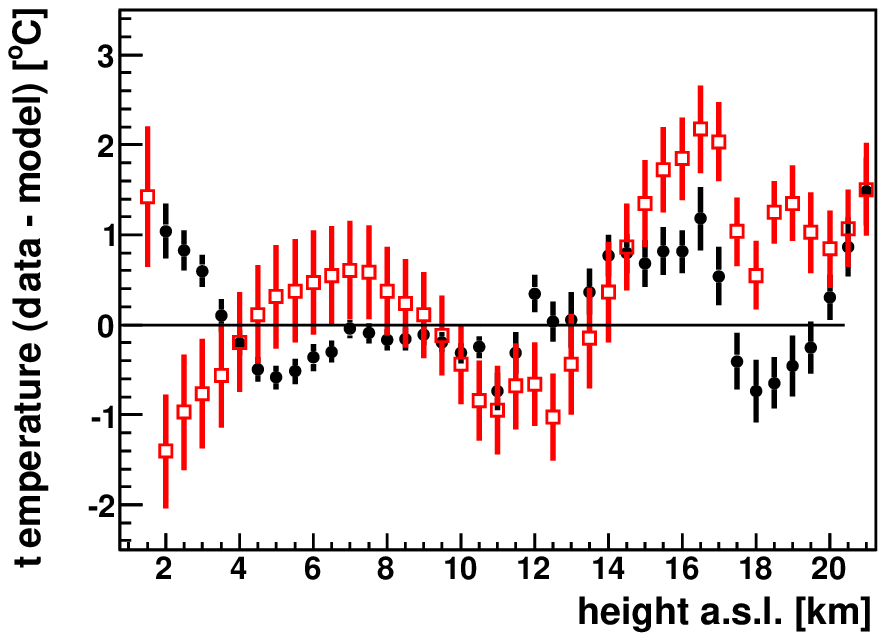}
  \end{minipage}
  \hfill
  \begin{minipage}[t]{.49\textwidth}
    \centering
    \includegraphics*[width=.99\linewidth,clip]{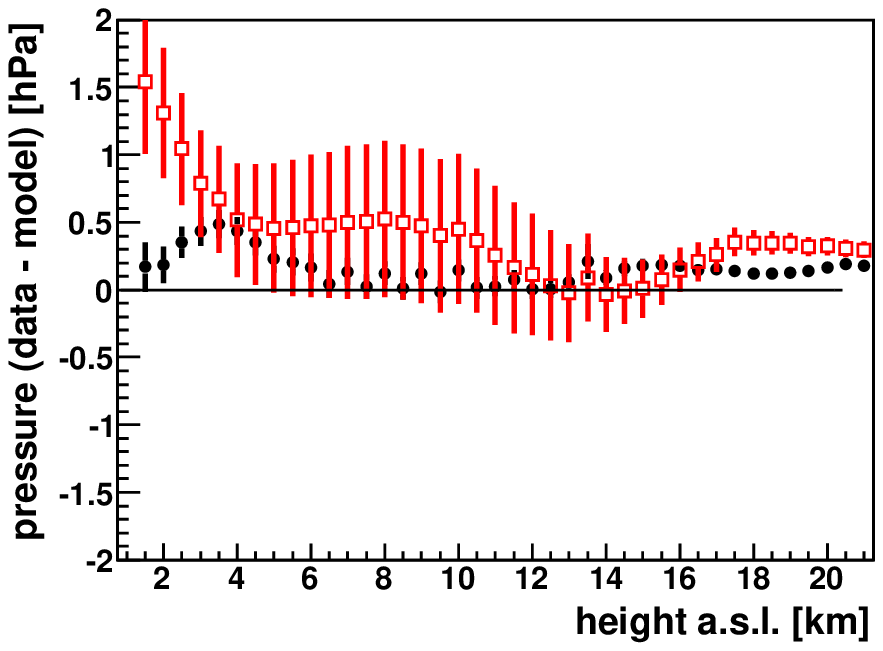}
  \end{minipage}
  \begin{minipage}[t]{.49\textwidth}
    \centering
    \includegraphics*[width=.99\linewidth,clip]{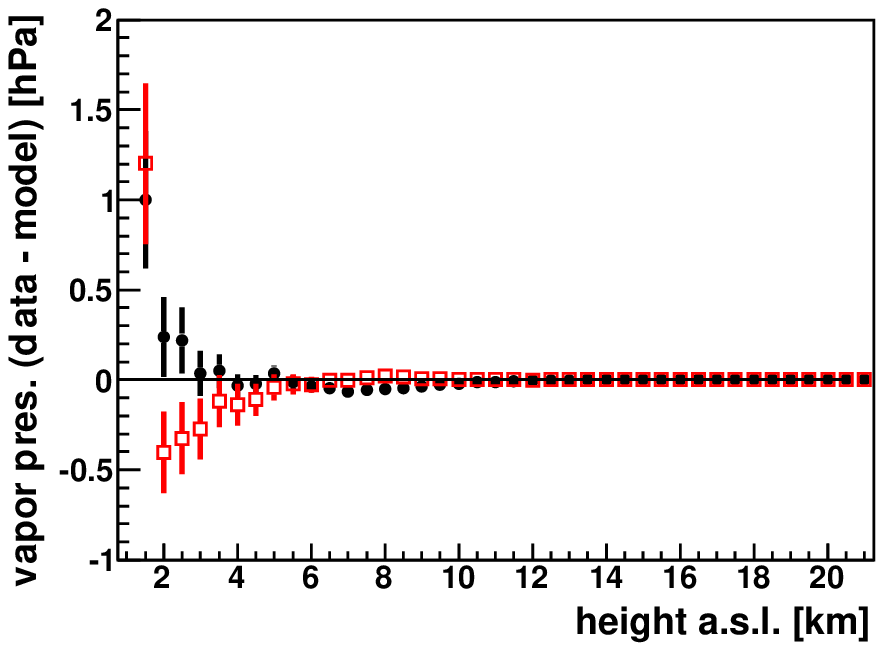}
  \end{minipage}
  \hfill
  \begin{minipage}[t]{.49\textwidth}
    \centering
    \includegraphics*[width=.99\linewidth,clip]{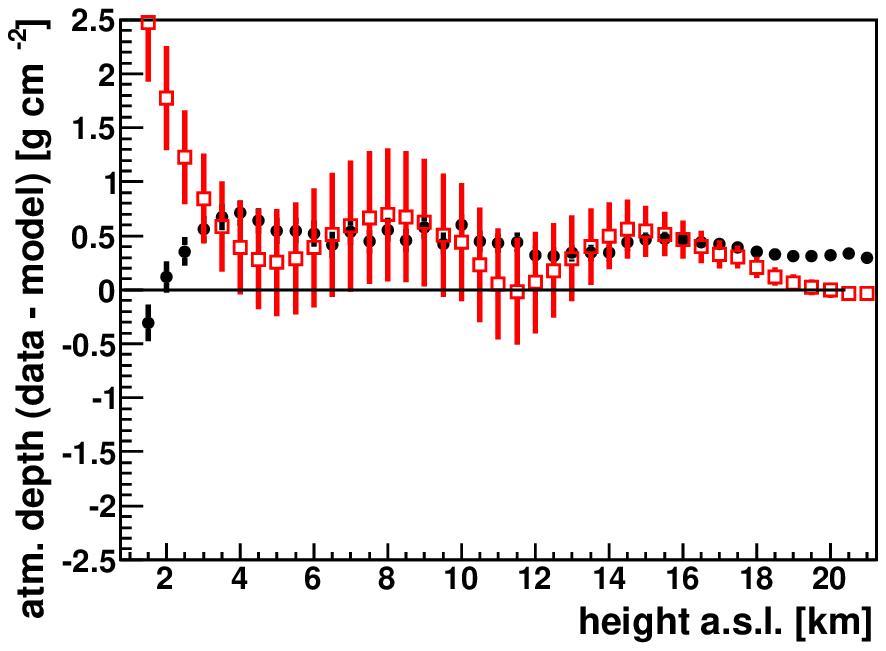}
  \end{minipage}
 \vspace{-12pt}
  \caption{\label{fig:SondeVsGDAS_nMMM_2009}
    Difference of measured radiosonde data and GDAS models (black dots) and
    monthly mean profiles (red squares) versus height for all radiosondes
    performed in 2009 and 2010.
  }
\end{figure}

In Fig.~\ref{fig:SondeVsGDAS_nMMM}, the GDAS data are compared with the measured
radiosonde data between 2005 and 2008. The comparison between the radiosonde
data and the monthly mean profiles is also shown. The monthly mean profiles are
averaged from the sounding data until end of 2008, thus, they fit local sounding
data of this period very well, see red squares in
Fig.~\ref{fig:SondeVsGDAS_nMMM}.  The error bars denote the RMS of the
differences at each height. The wave-like shape in the difference graph for the
atmospheric depth $X$ is driven by the piecewise parameterization of $X$ in the
nMMM.  Above around 5~km, the GDAS data deviate only marginally from the
measured data (black dots in Fig.~\ref{fig:SondeVsGDAS_nMMM}). Closer to the
ground, slightly larger differences become apparent. The measured temperature is
consistently higher than the model temperature. However, this might be caused by
two problems acting in the same direction. For our local radio soundings, the
temperature sensor might be quite often not properly acclimatized to outside
conditions but is affected by the inside temperature of the Balloon Launching
Station. It takes some minutes to overcome this effect during which time the
weather balloon is already launched. When compiling the nMMM, only temperature
data at 1600~m~a.s.l.\ and above were taken from the radio soundings. To
extrapolate to lower altitudes, a fit to these sounding data combined with data
from local weather stations at 1401~m, 1420~m, 1423~m, 1483~m, and 1719~m was
performed. However, data from the weather stations might also be influenced by
the direct surface conditions beneath, since they are not standardized
meteorological stations. It is also possible that the GDAS model does not
adequately describe the heated surface of the elevated plain of the Pampa
Amarilla and tends to assume free-atmosphere conditions. The pressure data of
both models are in good agreement. The water vapor pressure fits almost
perfectly, although the model values close to ground are both too low compared
to real data which can be traced back to the difficult handling of humidity in
general. In particular, the pressure differences propagate into the atmospheric
depth where we see deviations from the measured data on the same scale as for
the monthly mean profiles.

In the comparison displayed in Fig.~\ref{fig:SondeVsGDAS_nMMM_2009}, only
radiosonde data from 2009 and 2010 are used in order to illustrate the strength
of the GDAS model data. The nMMM are completely independent of this set of radio
soundings. The most obvious changes compared with
Fig.~\ref{fig:SondeVsGDAS_nMMM} are the worse descriptions of actual radio
soundings with monthly mean profiles. Even though the uncertainties become
larger, the nMMM do not describe the conditions measured during the years 2009
and 2010 very well. In contrast, the GDAS data can represent the local
conditions much better and the intrinsic uncertainty is consistently small.

\begin{figure}[htb]
  \begin{minipage}[t]{.49\textwidth}
    \centering
    \includegraphics*[width=.99\linewidth,clip]{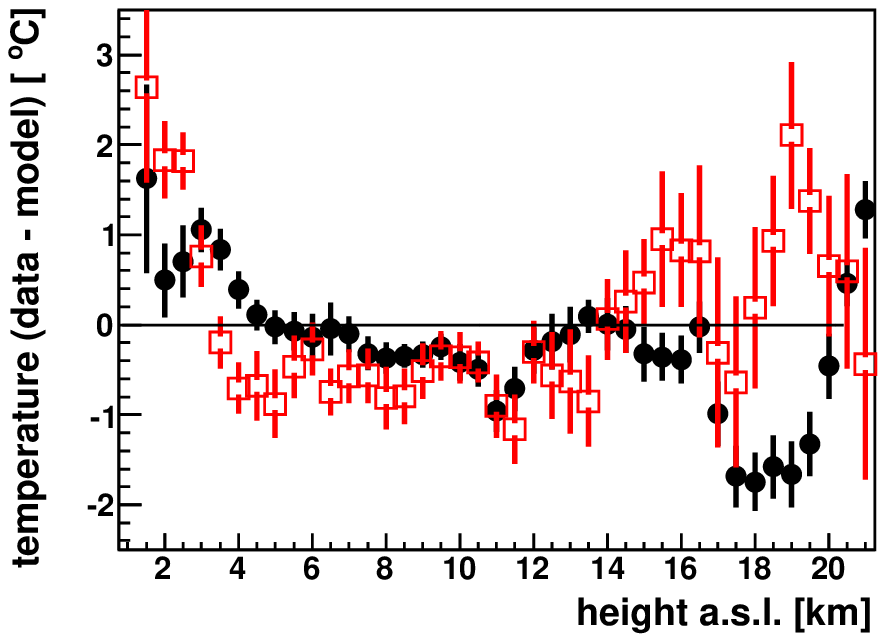}
  \end{minipage}
  \hfill
  \begin{minipage}[t]{.49\textwidth}
    \centering
    \includegraphics*[width=.99\linewidth,clip]{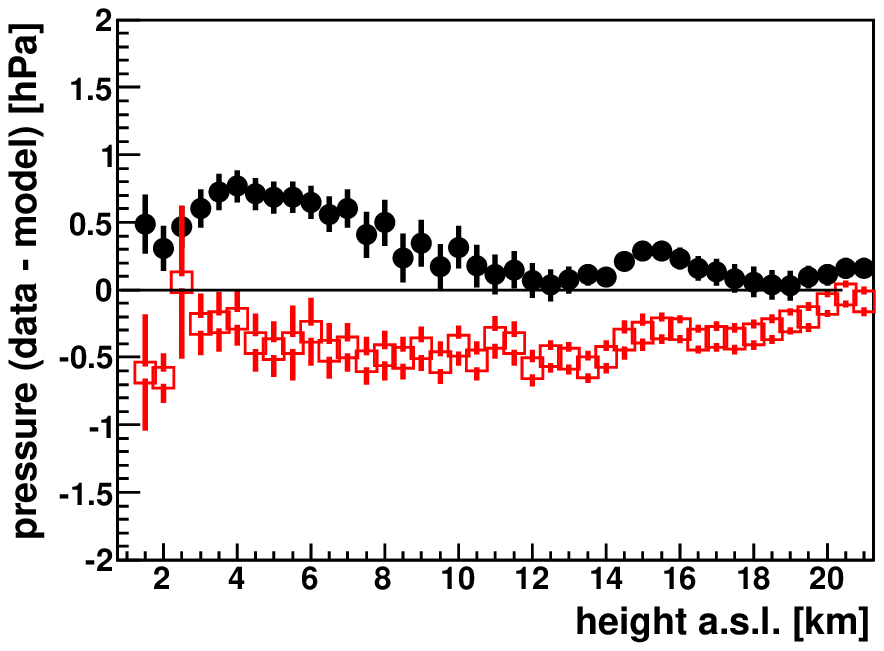}
  \end{minipage}
  \begin{minipage}[t]{.49\textwidth}
    \centering
    \includegraphics*[width=.99\linewidth,clip]{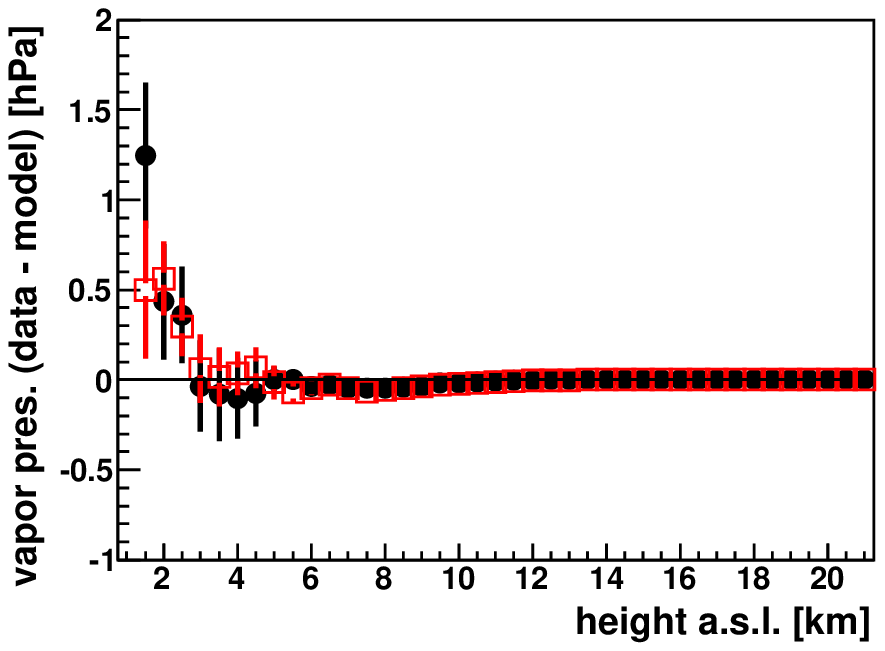}
  \end{minipage}
  \hfill
  \begin{minipage}[t]{.49\textwidth}
    \centering
    \includegraphics*[width=.99\linewidth,clip]{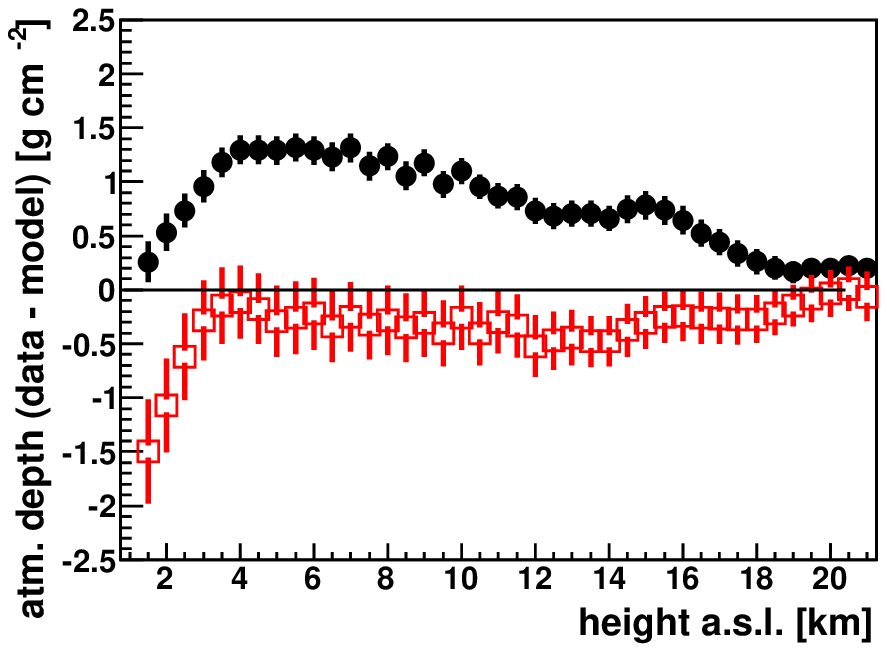}
  \end{minipage}
  \caption{\label{fig:SondeVsGDAS_short_long}
    Difference of measured radiosonde data and GDAS models for short (black
    dots) and long ascents (red squares) versus height.
  }
  \hfill
\end{figure}

In the following, the non-perfect agreement between GDAS data and local sounding
data below approximate 5~km~a.s.l.\ is investigated in more detail. This effect
close to ground could be caused by the nearby Andes and their influence on the
climate above the array. GDAS was developed for global atmospheric predictions
and therefore could be inadequate for very local atmospheric conditions. Under
normal circumstances, the wind carries weather balloons launched at the Pampa
Amarilla north-east or east with a horizontal displacement of about 100 to
150~km for ascents up to about 20~km~a.s.l. In some cases however, the balloon
just ascends with only a small horizontal drift in any direction, whereas the
opposite extreme cases are horizontal displacements of more than 200~km. For
this study, 25~short and localized ascents and 18~launches with very long
balloon paths were selected manually. The differences between GDAS data and the
measured data for both groups are shown in
Fig.~\ref{fig:SondeVsGDAS_short_long}. The pressure data from GDAS do not
describe these extreme conditions at the Auger Observatory as well as those seen
under normal circumstances. For extremely short soundings, the local
measurements reveal a high pressure area close to ground. This typically quite
local effect is not reproduced by GDAS but for higher altitudes, the model data
fit very well again. High pressure zones often indicate a stable atmospheric
layering, so conditions change only on long-term scales. This could cause the
good description of temperature profiles by GDAS. For the other extreme case,
the launches with very long balloon paths, the local pressure data indicate a
low pressure area, accompanied by significant winds. The conditions are
dominated by turbulences, indicated by short-term and small-scale temperature
variations.  Thus, the GDAS model data do not fit the local measurements well.
Overall, it is rather difficult to say if the topology of the Andes is the
source of uncertainty or if it is also an effect of quite extreme weather
conditions which might be induced by the structure of the Andes in the vicinity
of the Pampa Amarilla. Moreover, both groups of investigated ascents are
extremes and do not describe the usual conditions.  Less than 20\% of the
ascents launched at the \pao fall into either category.

Finally, possible inconsistencies between local measurements and GDAS data close
to the surface are investigated by using weather station data. While radiosondes
may suffer from measurement uncertainties near the ground, the weather stations
are specifically designed to continuously measure the ground values.

\subsection{GDAS vs.\ Ground Weather Stations
\label{sec:GDASvsWS}}

Five weather stations continuously monitor atmospheric values close to the
ground, at about 2 to 4~m above surface level. Four are located at the FD
stations, and one was set up near the center of the array at the Central Laser
Facility (CLF).  To make sure that the GDAS data describe the conditions at the
ground reasonably well, the values provided by the GDAS data set are compared to
all available weather station data. The GDAS data are interpolated at the
height of the weather station.

\begin{figure}[p]
  \begin{minipage}[t]{.56\textwidth}
    \centering
    \includegraphics*[width=.99\linewidth,clip]{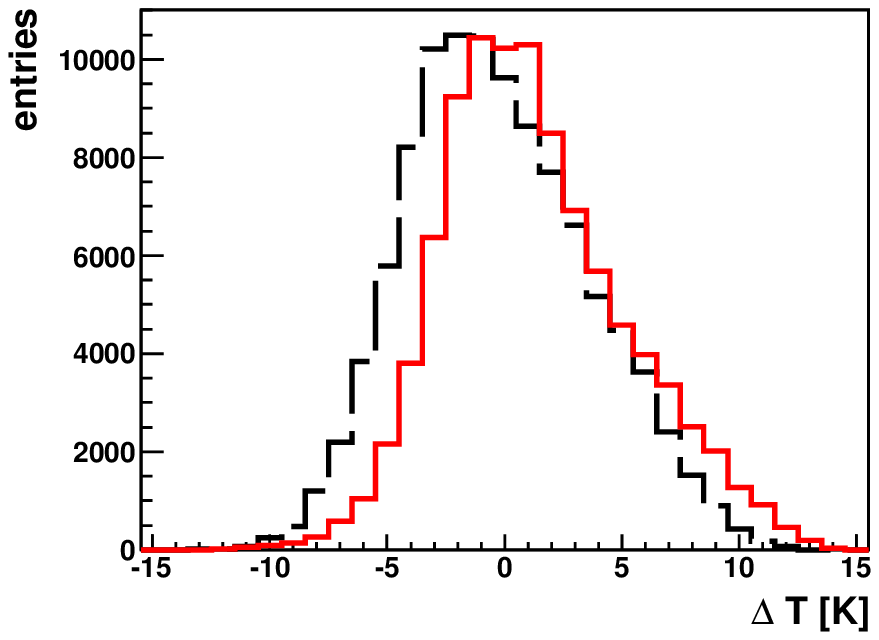}
  \end{minipage}
  \hfill
  \begin{minipage}[t]{.435\textwidth}
    \centering
    \includegraphics*[width=.99\linewidth,clip]{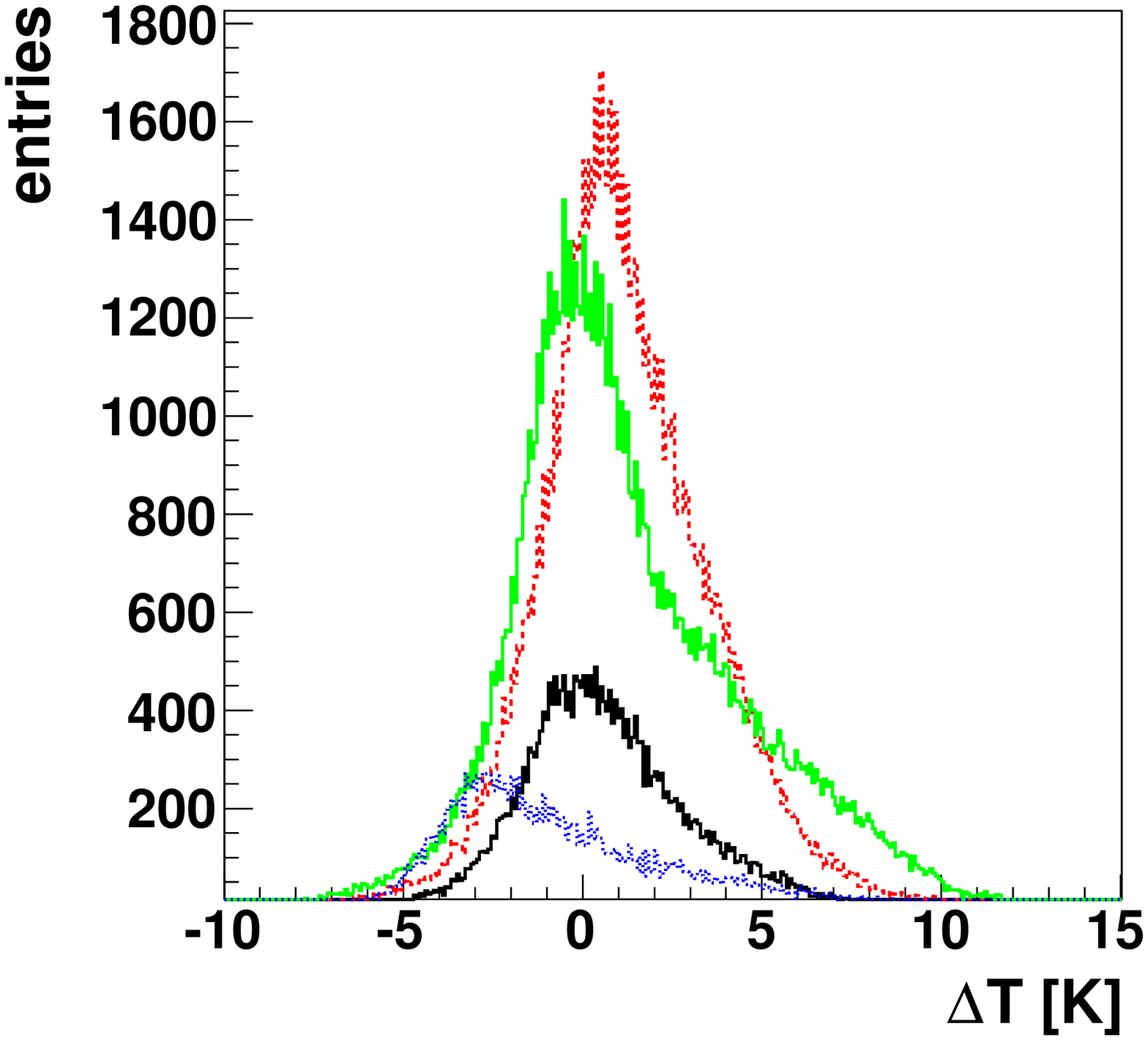}
  \end{minipage}
  \begin{minipage}[t]{.56\textwidth}
    \centering
    \includegraphics*[width=.99\linewidth,clip]{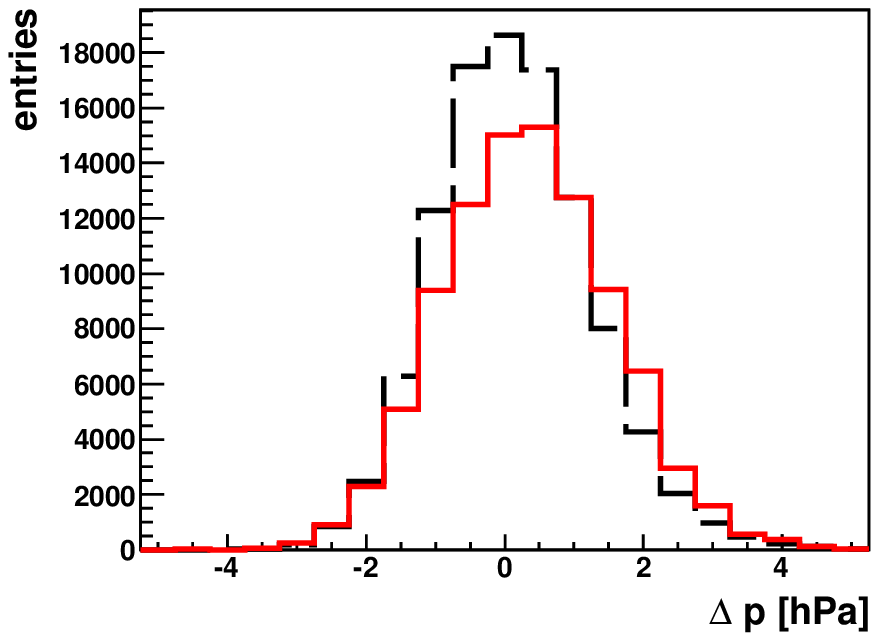}
  \end{minipage}
  \hfill
  \begin{minipage}[t]{.435\textwidth}
    \centering
    \includegraphics*[width=.99\linewidth,clip]{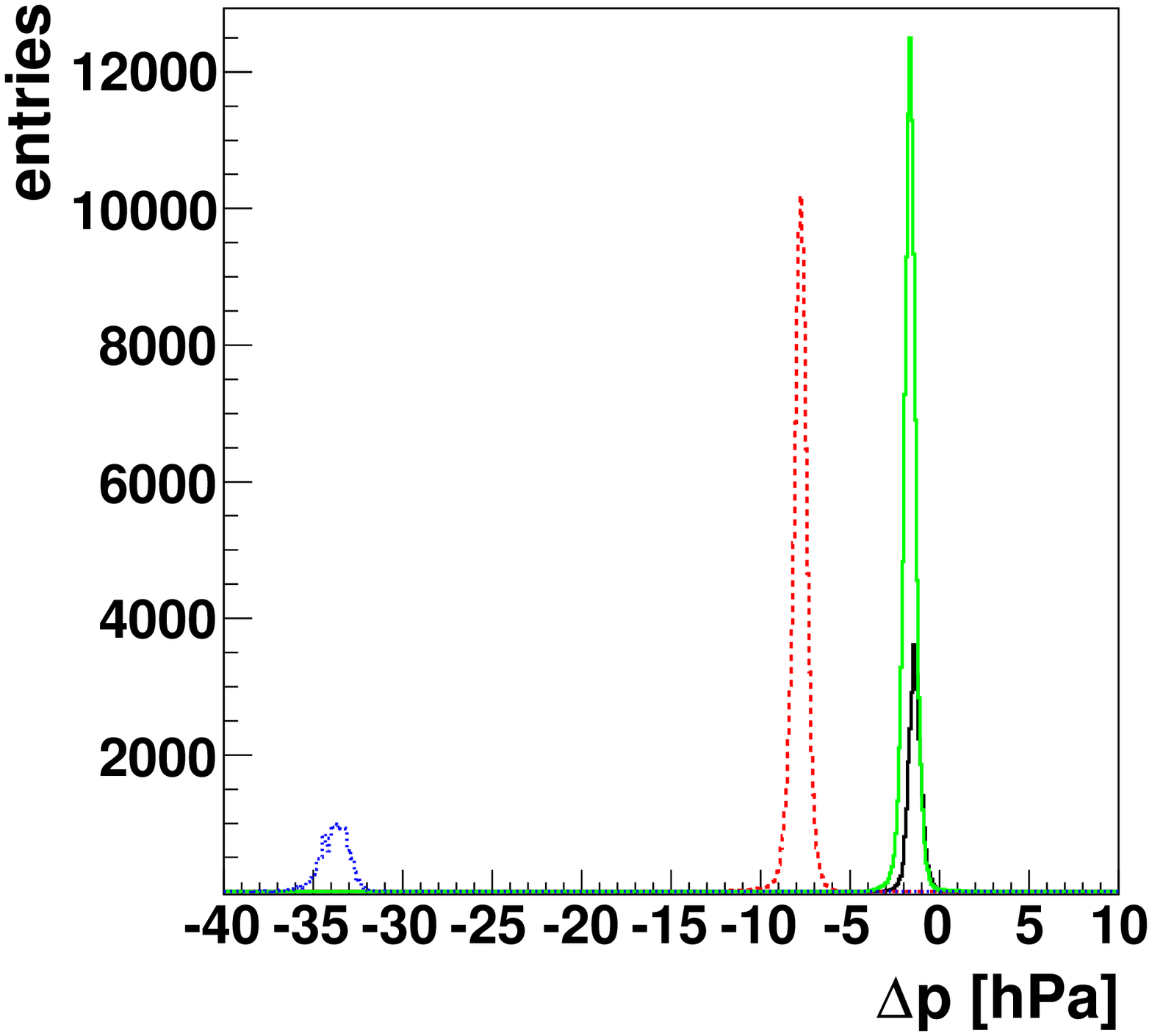}
  \end{minipage}
  \begin{minipage}[t]{.56\textwidth}
    \centering
    \includegraphics*[width=.99\linewidth,clip]{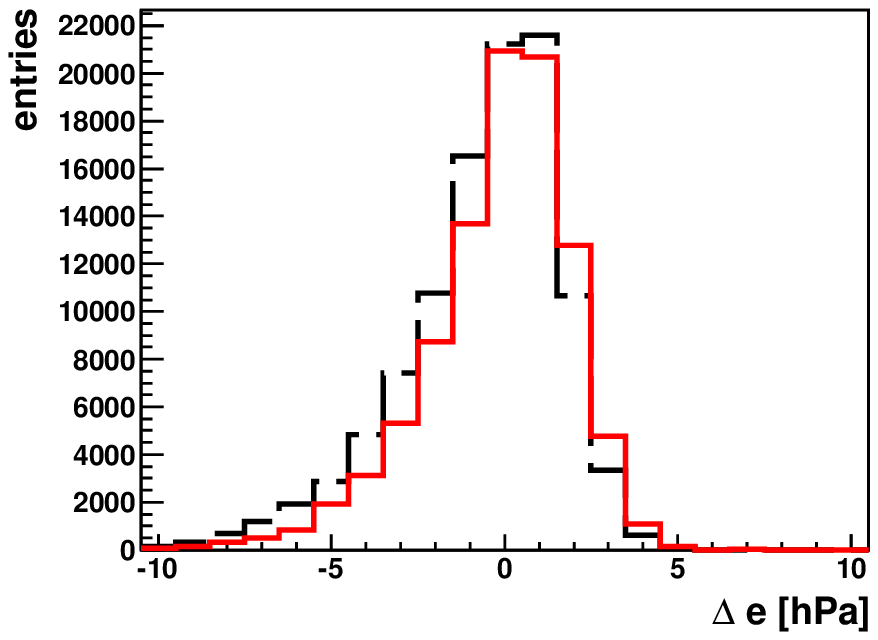}
  \end{minipage}
  \hfill
  \begin{minipage}[t]{.435\textwidth}
    \centering
    \includegraphics*[width=.99\linewidth,clip]{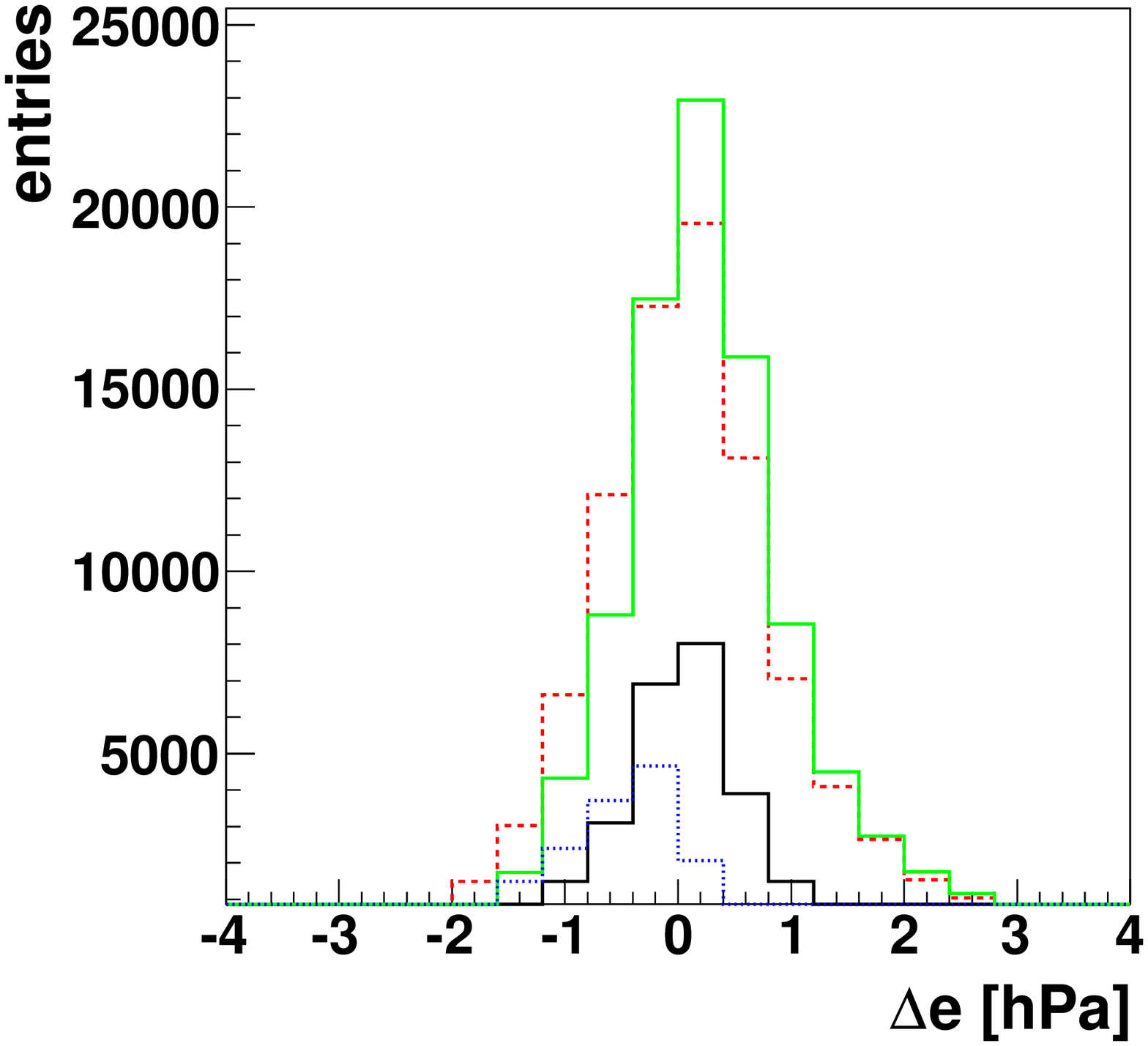}
  \end{minipage}
  \caption{\label{fig:WSvsGDAS}
    Difference of measured weather station data and GDAS models for all data
    from 2009. The mean and variance values for all histograms shown can be
    found in Table~\ref{tab:histo_values}. \newline On the left, the differences
    GDAS model minus measured data in temperature $T$, pressure $p$, and water
    vapor pressure $e$ are shown for the weather station at the CLF (solid red
    line) and the station at FD Loma Amarilla (dashed black line). \newline On
    the right, the differences between the data of the individual stations are
    shown.  The solid black line represents the difference between data measured
    with the Los Leones (LL) weather station and the data from the CLF weather
    station. The solid green line corresponds to Los Morados (LM) minus CLF, the
    dashed red line is for Loma Amarilla (LA) minus CLF and the dotted blue line
    indicates Coihueco (CO) data minus CLF data.
  }
  \hfill
\end{figure}

In the histograms on the left of Fig.~\ref{fig:WSvsGDAS}, the differences
between measured weather station data and GDAS model data are shown for the
weather stations close to the CLF and FD Loma Amarilla. All data measured in
2009 were used. Temperature, pressure, and water vapor pressure agree very well.
Details of the histograms are listed in Tab.~\ref{tab:histo_values}.

\begin{table}[t]
  \begin{center}
    \caption{\label{tab:histo_values}
      \small{Mean and variance values for the histograms shown in
      Fig.~\ref{fig:WSvsGDAS}. Values for the pressure were not corrected for
      the height difference of the stations, but they are consistent with those
      differences.  Applying the USStdA, the pressure difference due to the
      different altitudes of the weather stations are $-$2.0~hPa (LL),
      $-$8.6~hPa (LA), $-$2.3~hPa (LM), and $-$32.8~hPa (CO).}
    }
    \vspace{10pt}
    \begin{tabular}{crrrcrrr}
     \hline
      & \multicolumn{3}{c}{Mean} &  & \multicolumn{3}{c}{RMS} \\
      \cline{2-4} \cline{6-8}
      & $T$ [K] & $p$ [hPa] & $e$ [hPa] &
      & $T$ [K] & $p$ [hPa] & $e$ [hPa] \\
      \hline
      $x_{\rm{WS}} - x_{\rm{GDAS}}$ & & & & & & & \\
      CLF &    1.3 &   0.4 & $-$0.2 & & 3.9 & 1.2 & 2.1 \\
      LA  & $-$0.3 &   0.2 & $-$0.7 & & 3.9 & 1.1 & 2.3 \\
      \hline
      $x_{\rm{WS}} - x_{\rm{CLF}}$  & & & & & & & \\
      LL  &    0.6 &  $-$1.4 & $-$0.2 & & 2.4 & 0.4 & 1.2 \\
      LA  &    1.2 &  $-$7.8 &    0.2 & & 2.5 & 0.5 & 1.3 \\
      LM  &    1.2 &  $-$1.7 &    0.3 & & 3.2 & 0.4 & 1.1 \\
      CO  & $-$0.6 & $-$33.9 & $-$0.9 & & 3.2 & 0.8 & 1.5 \\
      \hline
    \end{tabular}
  \end{center}
\end{table}

On the right side of Fig.~\ref{fig:WSvsGDAS}, the differences of the data of the
individual weather stations are shown. The CLF weather station is close to the
middle of the array and was chosen as a reference. Values for the pressure were
not corrected for the height difference of the stations, but they are consistent
with the height differences of the stations. The mean and width of the CLF-LA
distribution is very similar to the distributions on the left for the GDAS data.
These two histograms are expected to be similar because of the vicinity of the
selected GDAS grid point and the FD building Loma Amarilla.  Overall, the
differences between the GDAS data and the weather station data are of the same
order as the difference in data of two different weather stations.  Only the
difficult predictability of water vapor in the atmosphere close to ground can be
seen again.

\begin{figure}[tbp]
  \begin{minipage}[t]{.49\textwidth}
    \centering
    \includegraphics*[width=.99\linewidth,clip]{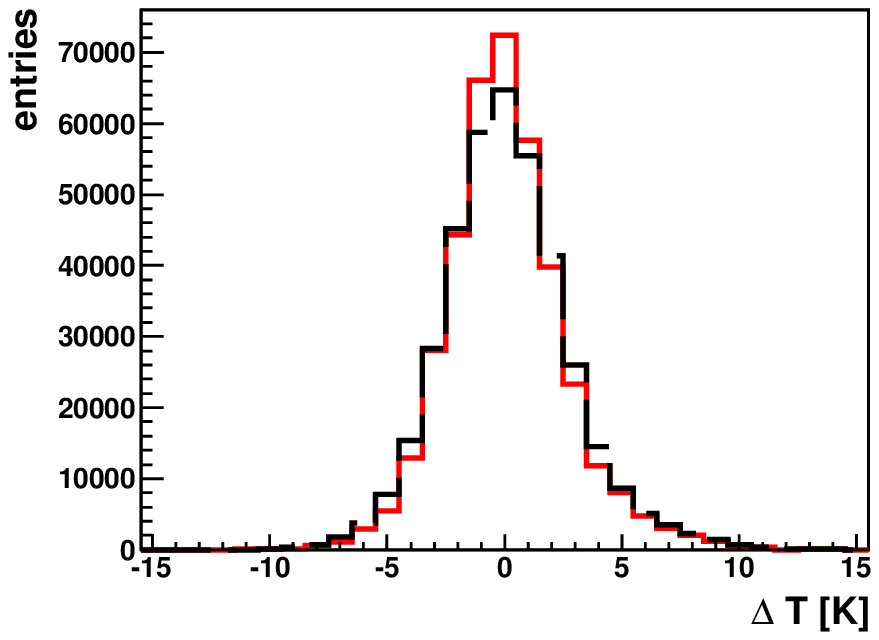}
  \end{minipage}
  \hfill
  \begin{minipage}[t]{.49\textwidth}
    \centering
    \includegraphics*[width=.99\linewidth,clip]{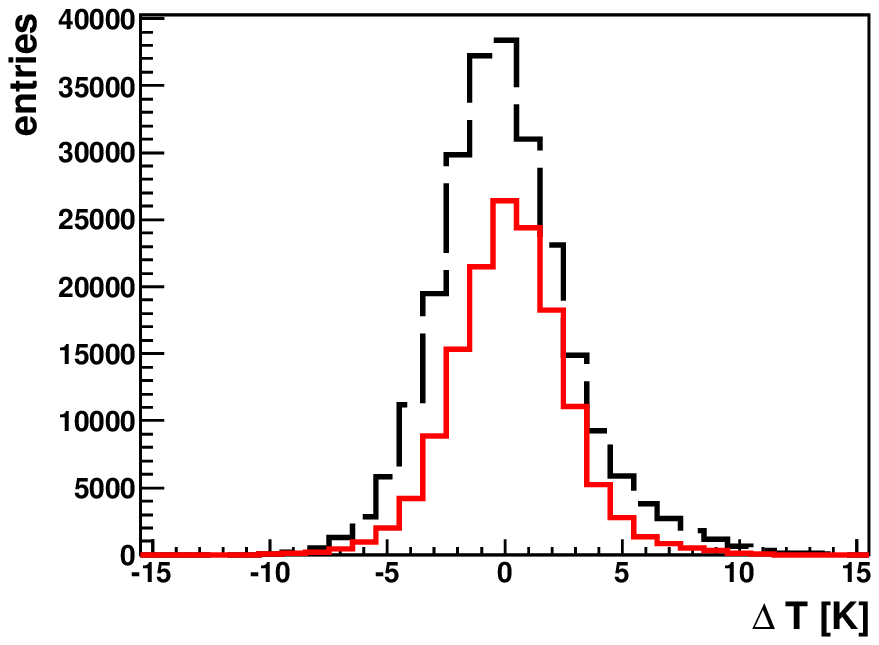}
  \end{minipage}
  \begin{minipage}[t]{.49\textwidth}
    \centering
    \includegraphics*[width=.99\linewidth,clip]{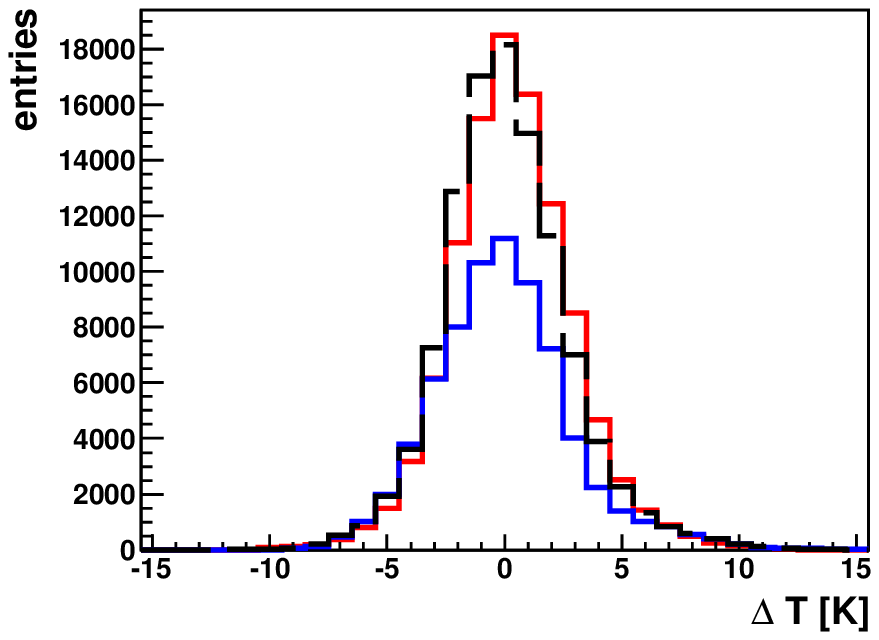}
  \end{minipage}
  \hfill
  \begin{minipage}[t]{.49\textwidth}
    \centering
    \includegraphics*[width=.99\linewidth,clip]{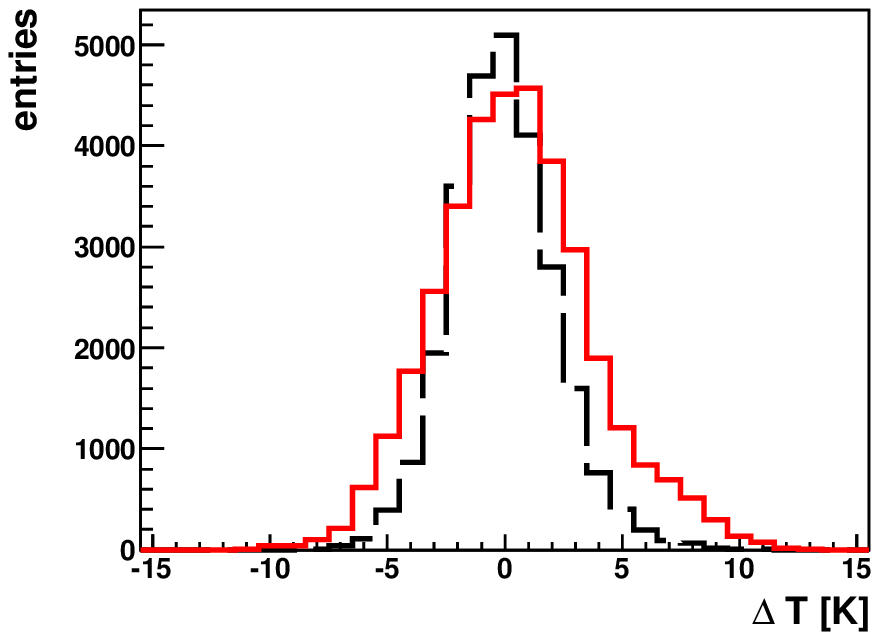}
  \end{minipage}
  \caption{\label{fig:WSvsGDAS_LM}
    The difference of GDAS minus weather station data in temperature is shown
    for the weather station at Los Morados. At the top left, the difference of
    all data measured every 5~minutes (dashed black) is compared to the
    difference only of data measured every three hours when new GDAS data are
    available (solid red, scaled by 36). At the top right, all data are split
    into daytime (dashed black) and nighttime (solid red) measurements. At the
    bottom left panel, all data are separated by years (2007 in solid blue, 2008
    in solid red, 2010 in dashed black).  At the bottom right, data are split
    into summer and winter, represented by one month each (January in dashed
    black, July in solid red).
  }
  \hfill
\end{figure}

\begin{figure}[tbp]
  \begin{minipage}[t]{.49\textwidth}
    \centering
    \includegraphics*[width=.99\linewidth,clip]{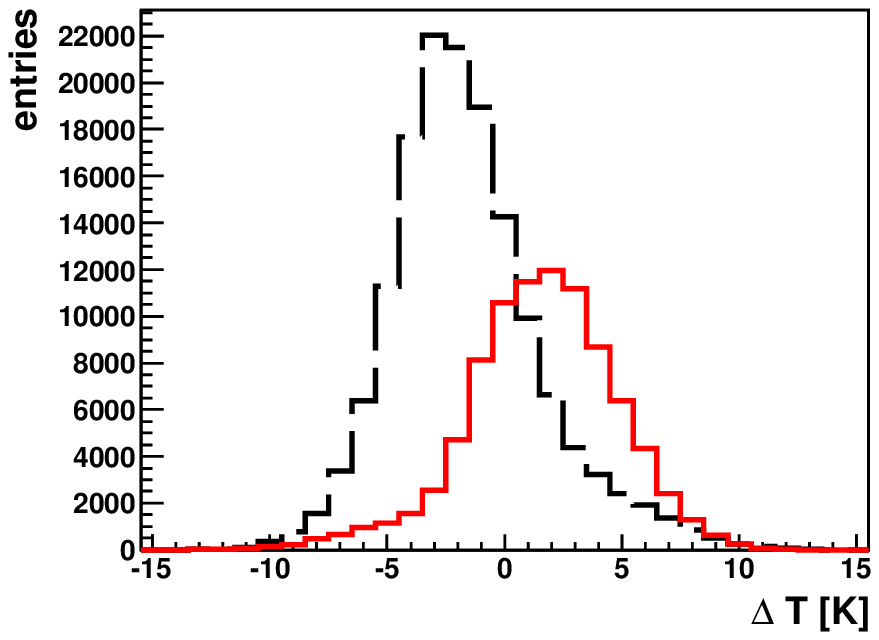}
  \end{minipage}
  \hfill
  \begin{minipage}[t]{.49\textwidth}
    \centering
    \includegraphics*[width=.99\linewidth,clip]{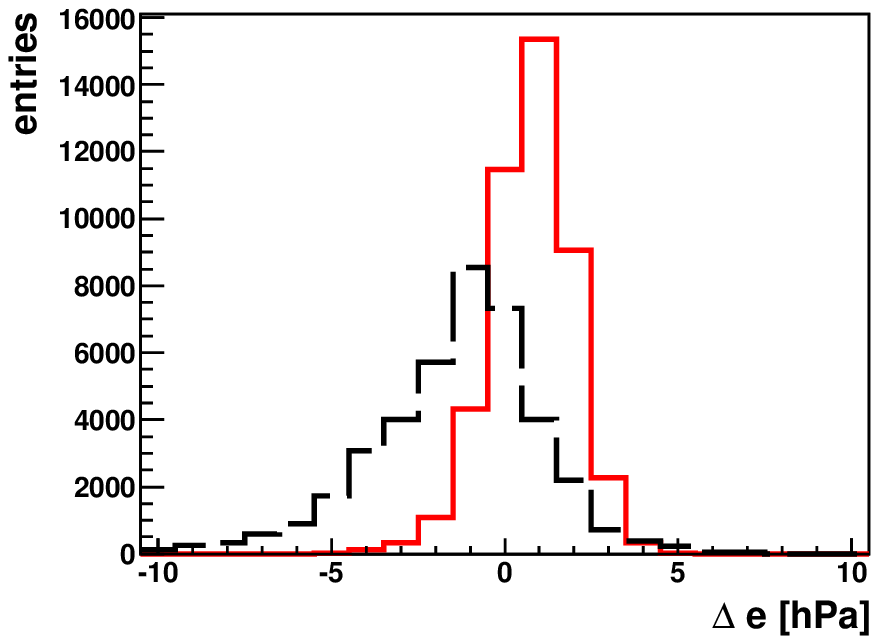}
  \end{minipage}
  \caption{\label{fig:WSvsGDAS_dev}
    The difference of GDAS minus weather station data in temperature is shown in
    the left panel for the station at Loma Amarilla, separated by time of day
    (nighttime in dashed black, daytime in solid red). In the right histogram,
    the water vapor pressure for the station at the CLF is shown separated by
    seasons (Summer in dashed black, winter in solid red).
  }
  \hfill
\end{figure}

In Fig.~\ref{fig:WSvsGDAS_LM}, the differences in temperature for the weather
station at Los Morados are analyzed in more detail. At the top left, all data
measured every 5~minutes are compared with GDAS data. Additionally, only station
data every three hours at which new GDAS data are available are shown, scaled by
a factor of 36. Both distributions show identical means of 0.1~K, as well as
similar RMS values of 2.7~K and 2.6~K for all data and for the 3-hourly data,
respectively. The atmospheric parameters at the Observatory are very stable
during 3-hour periods. At the top right, the data are split into day and night.
Night is defined as the UTC hours between 0 and 10.  Therefore, with 14 compared
to 10~hours of data, the daytime distribution contains more data. For Los
Morados, both distributions are close to each other, with a mean of 0.0~K and
RMS of 2.9~K during the day and a mean of 0.2~K and RMS of 2.4~K at night. A
small difference at night is noticeable, with the GDAS data giving higher
temperatures than the weather station. The distributions for different years
(bottom left) and different seasons (bottom right), represented by one month of
summer and one from winter, show no distinct features or differences. Similar
studies for pressure and water vapor pressure mostly yield similar results.
Nevertheless, two deviations from this general pattern are discussed in the
following.

In the left panel of Fig.~\ref{fig:WSvsGDAS_dev}, the temperature distributions
for the station at Loma Amarilla are shown, split into daytime (solid red) and
nighttime (dashed black) measurements. A clear separation of the two
distributions is found. The mean of the distributions is $-$1.6~K for daytime
and 1.6~K for nighttime, resulting in a difference of 3.2~K. This difference
might be due to the local environment in which the weather station is placed.
While some stations are far away from the FD buildings, others had to be mounted
closer to or on top of other facilities. Thus, a standardized meteorological
measurement can not be guaranteed.

On the right of Fig.~\ref{fig:WSvsGDAS_dev}, the water vapor pressure for
different seasons measured at the weather station at the CLF is shown. Clear
differences are apparent. In austral winter (July), the water vapor pressure is
very low, and the measurements agree better with GDAS data, the mean of the
distribution is 0.8~hPa with an RMS of 1.2~hPa. In summer (January), the
deviations are largest, where the mean difference drops to $-$1.4~hPa, and the
RMS doubles to 2.4~hPa. This indicates that the GDAS description of humidity is
not perfect, underestimating the humidity in summer. However, the water vapor
pressure calculation strongly depends on the temperature, so differences in
temperature due to local effects of the surroundings of the station also affect
this comparison.

Apart from the differences that were seen between radiosonde data and GDAS data
near the surface in the previous section, the comparison of GDAS data with
weather station data shows a very good agreement. We conclude that the GDAS data
describe the conditions at the Pierre Auger Observatory very well. Because of
their highly reliable availability and high frequency of data sets, GDAS data are
a suitable replacement for local radio soundings and also for the local monthly
models.

\subsection{GDAS vs.\ Radio Soundings at Other Locations
\label{sec:GDASvsRadioUS}}

In addition to the measurements in Argentina, similar radiosonde launches were
performed in south-east Colorado, United States.  27~weather balloons were
launched between September 2009 and December 2010 using identical radiosondes
and equipment. The comparisons with the GDAS data for this location show
differences of the same order as for the location of the Pierre Auger
Observatory.  This is remarkable because most of the global atmospheric models,
in particular those developed in Europe and North America, typically describe
the conditions at the northern hemisphere much better. This is due to the fact
that atmospheric measurements in South America and in general at the southern
hemisphere are sparse and accurate modeling of the atmosphere is predicated on
real data.

Comparisons of GDAS data and radio soundings at further locations would go
beyond the scope of this paper. Moreover, the radio soundings from south-east
Colorado are independent data while other available data are from radio sounding
databases which are part of the global meteorological network used for the
creation of GDAS.

\section{Air Shower Reconstruction\label{sec:reco}}

To study the effects caused by using the GDAS data, all air shower data from the
Auger Observatory collected between June~1, 2005 and end of 2010 are used in a
reconstruction analysis using the \offline software framework of the Pierre
Auger Observatory~\cite{Argiro:2007qg}. The change of atmosphere description
will mainly affect the reconstruction of the fluorescence data, c.f.\
Sec.~\ref{sec:impact}.  Therefore, we concentrate on this part in the following.
It is known that varying atmospheric conditions alter the fluorescence light
production and transmission~\cite{Abraham:2010}. The transmittance of the actual
atmosphere is regularly measured during FD shifts and made available for air
shower reconstructions via databases.  The light production has to be calculated
analytically during the reconstruction procedure. Its strong
atmosphere-dependence as described in Sec.~\ref{sec:impact} is applied in the
air shower reconstruction analysis.

\subsection{Data Reconstruction}

The following analysis is based on three sets of reconstructions. The first set,
\textsf{FY}, is the until recently standard reconstruction of the Pierre Auger
Observatory. The fluorescence yield is calculated with its atmosphere-dependence
as described in \cite{Ave:2007}, along with the monthly mean profiles (nMMM)
obtained for the site of the Auger Observatory.  For the second set,
\textsf{FY$_{\rm mod}$}, all currently known atmospheric effects in the
fluorescence calculation are taken into account.  Together with the standard
atmosphere-dependence, the temperature-dependent collisional cross sections and
humidity quenching are included (\cite{Arqueros:2008} and references therein).
Parameterizations for these two effects are taken from AIRFLY~\cite{Ave:2008}
and later conference contributions by the AIRFLY collaboration. Again, the nMMM
are used.  The third set, \textsf{FY$_{\rm mod}^{\rm GDAS}$}, also explores the
efficiency of the full atmosphere-dependent fluorescence description, but here
the atmospheric nMMM are exchanged with the new 3-hourly GDAS data.

\begin{figure}[!t]
  \begin{minipage}[t]{.49\textwidth}
    \centering
    \includegraphics*[width=.99\linewidth,clip]{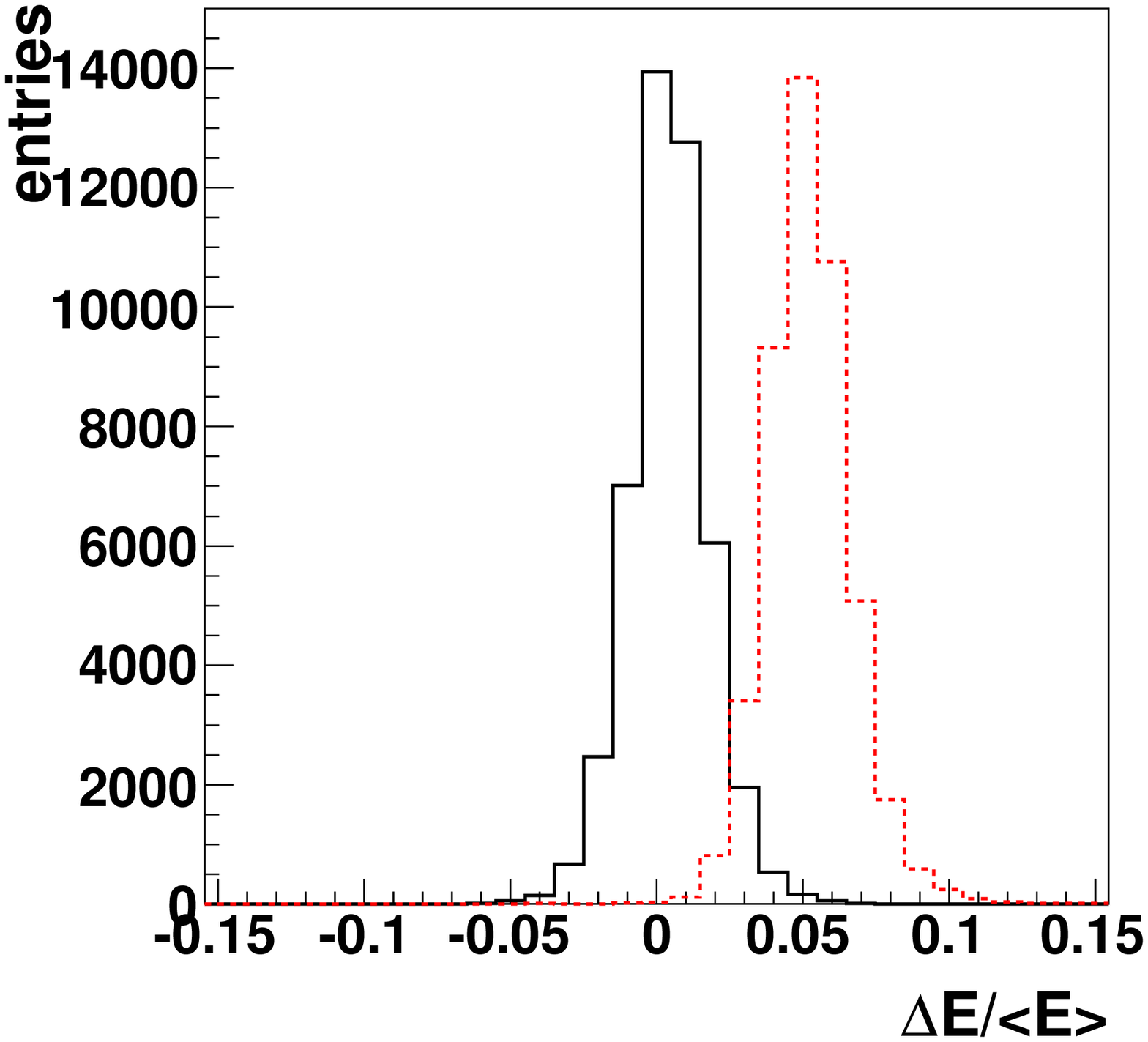}
  \end{minipage}
  \hfill
  \begin{minipage}[t]{.49\textwidth}
    \centering
    \includegraphics*[width=.99\linewidth,clip]{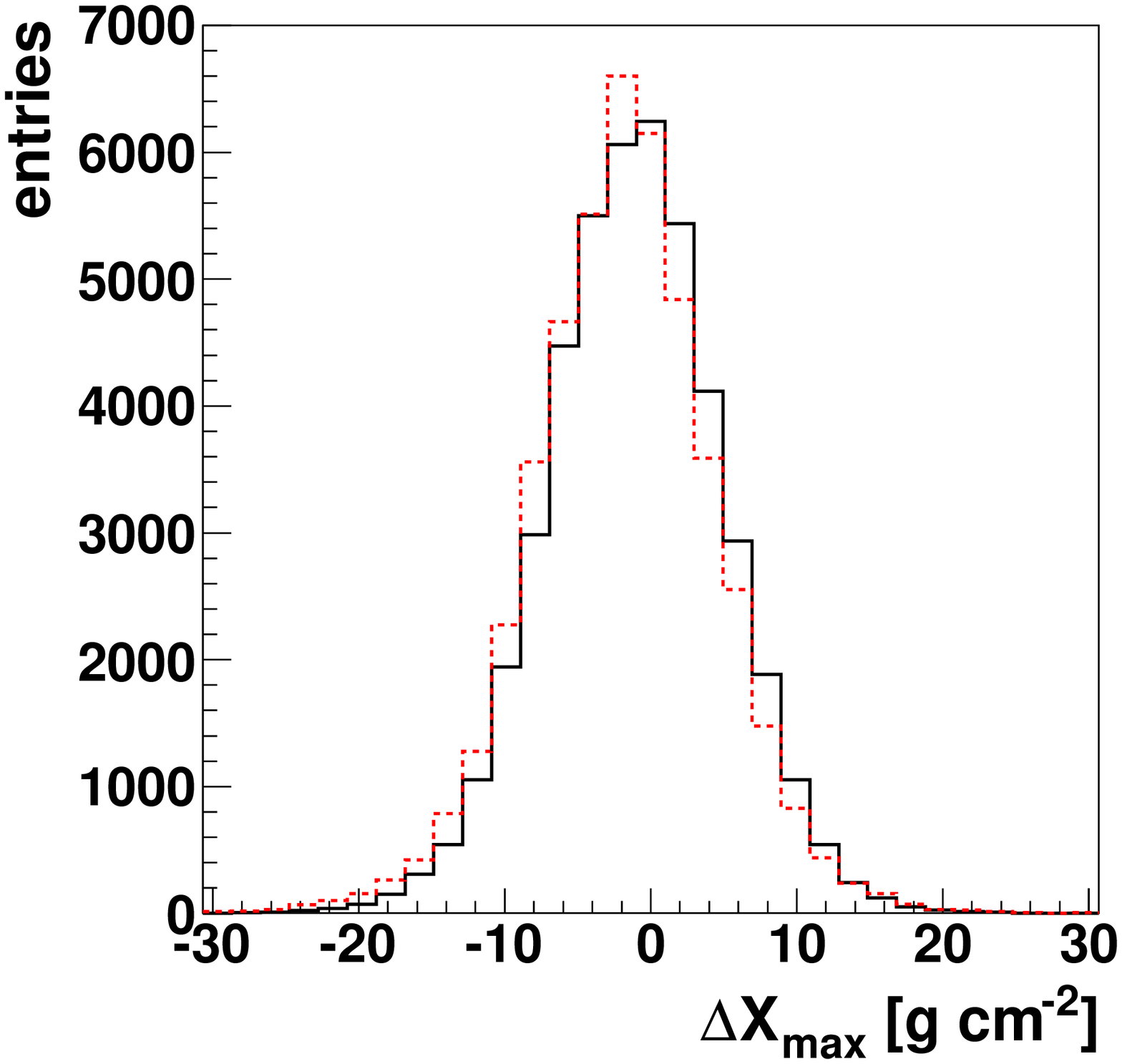}
  \end{minipage}
  \begin{minipage}[t]{.49\textwidth}
    \centering
    \includegraphics*[width=.99\linewidth,clip]{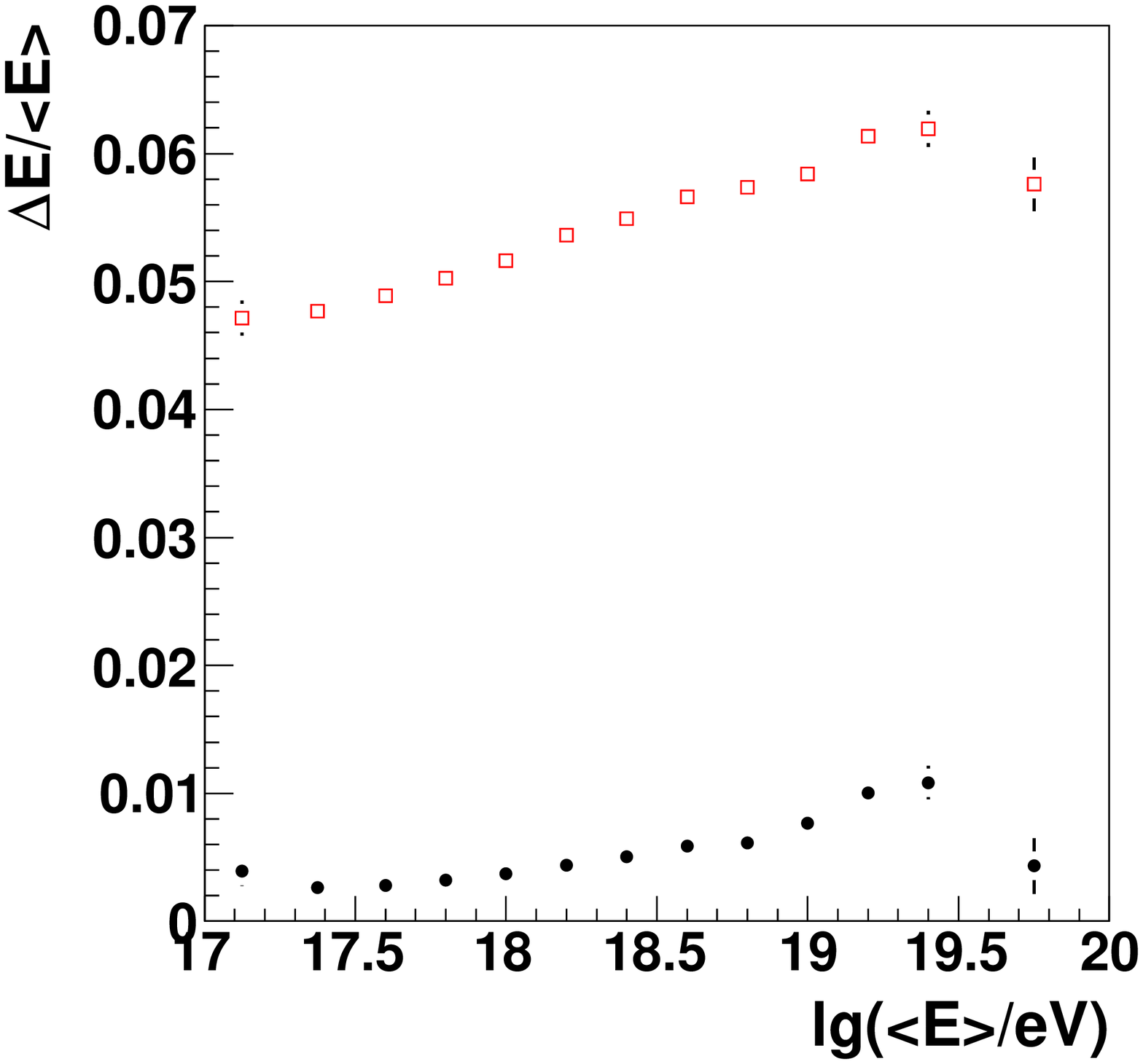}
  \end{minipage}
  \hfill
  \begin{minipage}[t]{.49\textwidth}
    \centering
    \includegraphics*[width=.99\linewidth,clip]{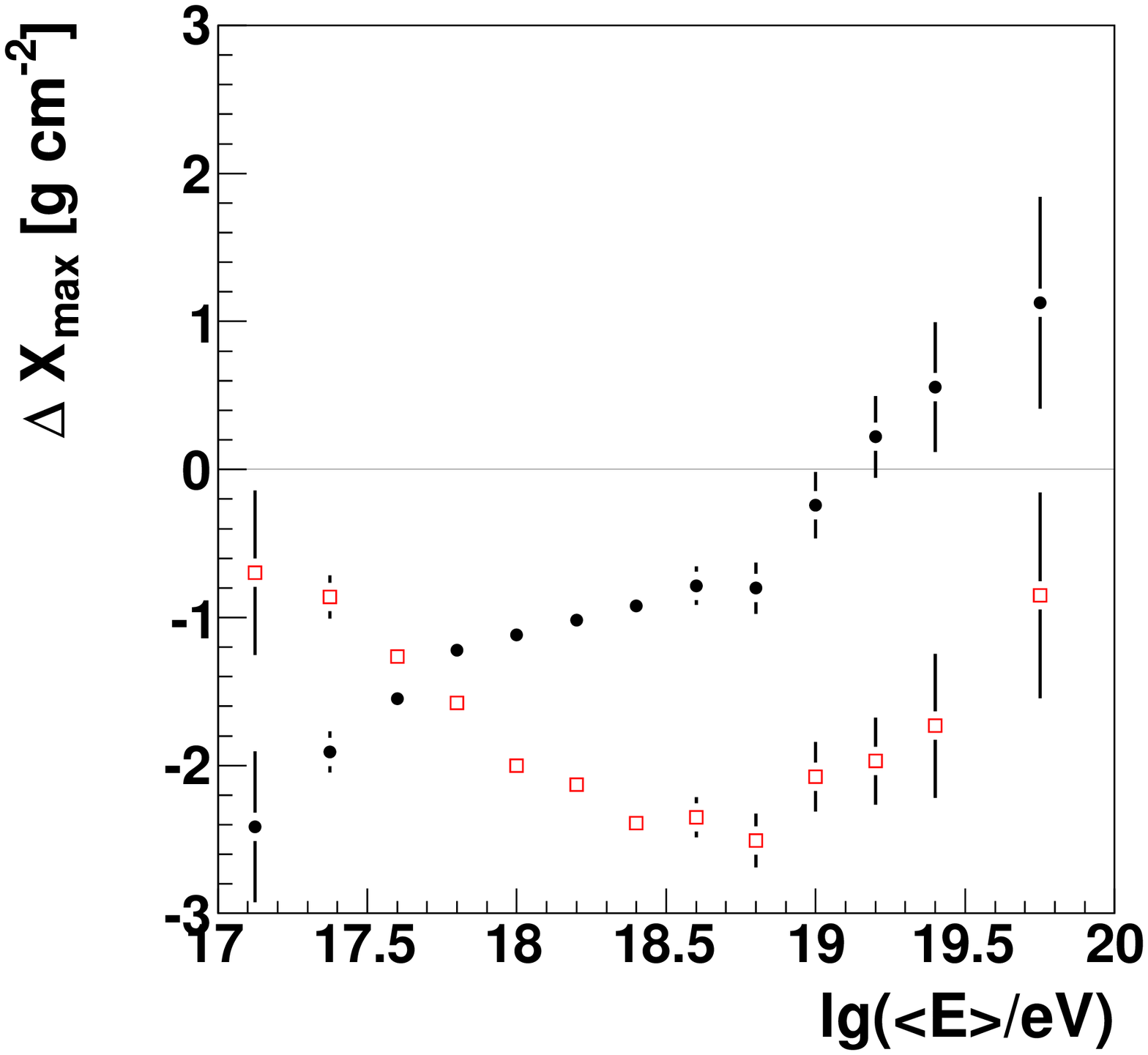}
  \end{minipage}
  \begin{minipage}[t]{.49\textwidth}
    \centering
    \includegraphics*[width=.99\linewidth,clip]{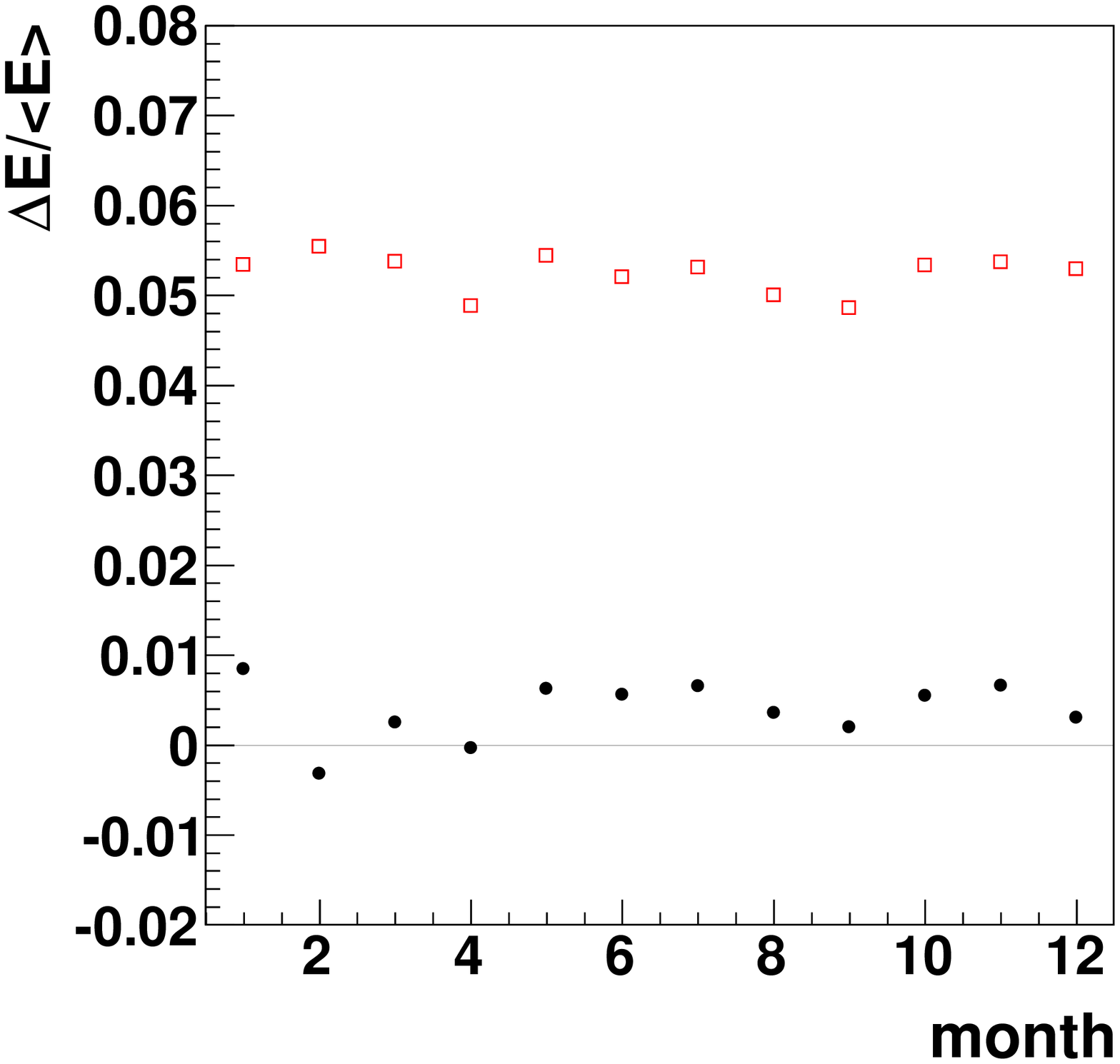}
  \end{minipage}
  \hfill
  \begin{minipage}[t]{.49\textwidth}
    \centering
    \includegraphics*[width=.99\linewidth,clip]{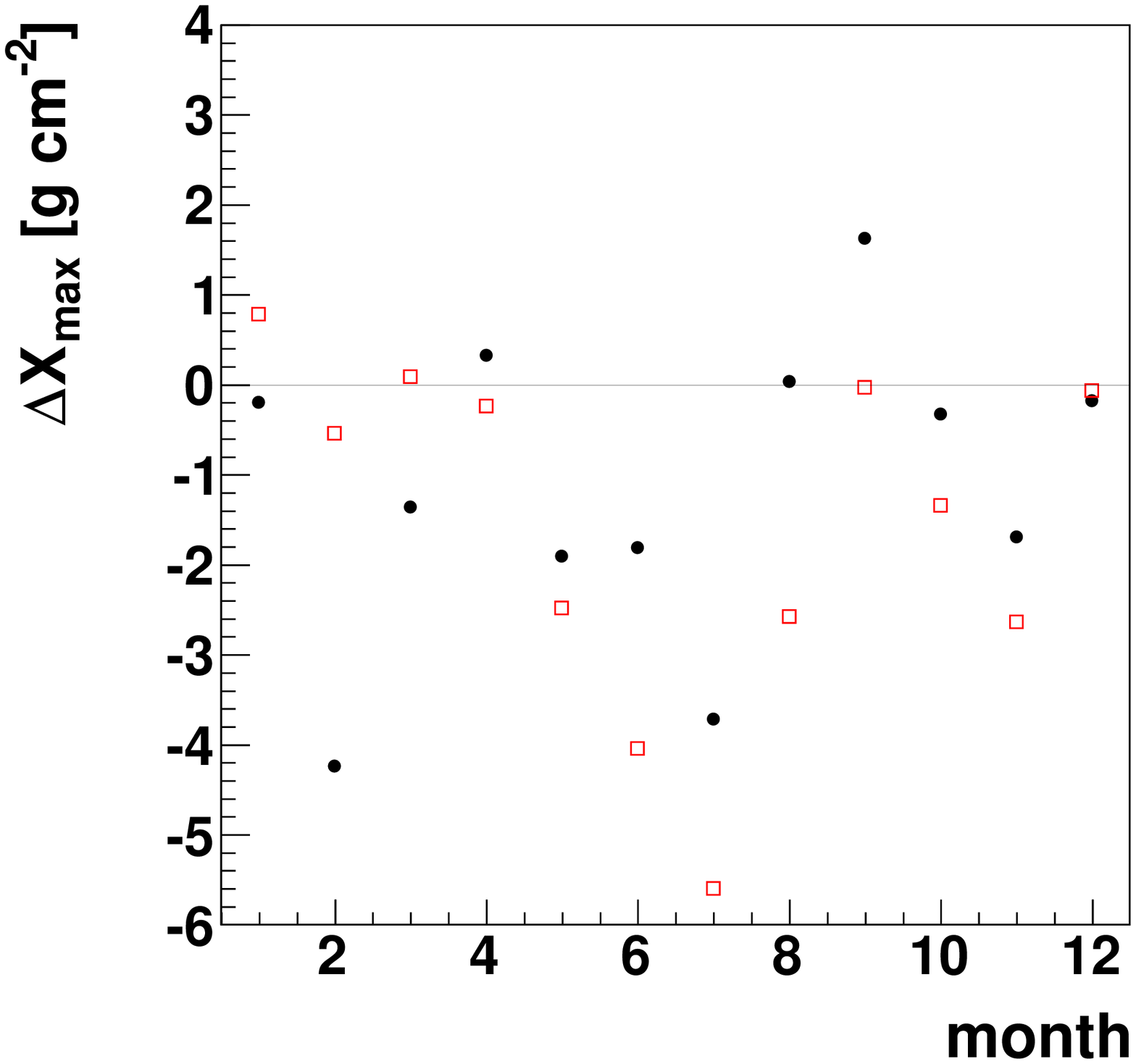}
  \end{minipage}
  \caption{\label{fig:delta}
    Difference of reconstructed $E$ (left) and \xmax (right). Top row: binned
    differences. Middle row: differences as a function of energy. Bottom row:
    differences as a function of month. Black solid line or filled dots for
    \textsf{FY$_{\rm mod}^{\rm GDAS}$} minus \textsf{FY$_{\rm mod}$}, and red
    dashed line or open squares for \textsf{FY$_{\rm mod}^{\rm GDAS}$} minus
    \textsf{FY}.
  }
\end{figure}

Comparing the reconstruction sets with each other, the variation of the
reconstructed primary energy $E$ of air showers and the position of shower
maximum \xmax can be determined, see Fig.~\ref{fig:delta}. In the two upper
figures, the binned difference of $E$ and \xmax is displayed, and the
dependences on energy and month of these differences are shown in the figures in
the middle and bottom, respectively.

Using GDAS data for the reconstruction instead of nMMM affects the reconstructed
primary energy only slightly. The mean of the difference \textsf{FY$_{\rm
mod}^{\rm GDAS}$} minus \textsf{FY$_{\rm mod}$} is 0.4\% with an RMS of 1.4\%
(Fig.~\ref{fig:delta}, top left, solid black line). For the reconstructed \xmax,
only a small shift of $-$1.1~\gcmsq is found with an RMS of 6.0~\gcmsq
(Fig.~\ref{fig:delta}, top right, solid black line). However, comparing the full
atmosphere-dependent reconstruction \textsf{FY$_{\rm mod}^{\rm GDAS}$} with the
unmodified reconstruction \textsf{FY}, a clear shift in $E$ can be seen.  An
increase in $E$ by 5.2\% (RMS 1.5\%) and a decrease of \xmax by $-$1.9~\gcmsq
(RMS 6.3~\gcmsq) is found. These modified fluorescence settings are now used in
the reconstruction of the Pierre Auger Observatory, in conjunction with other
improvements to the procedure, see~\cite{Pesce:2011}.

The difference in reconstructed $E$ vs.\ mean $E$ (Fig.~\ref{fig:delta}, middle
left) reveals a small energy dependence, increasing towards higher energies.
The description of atmospheric conditions close to the ground is very difficult
in monthly mean profiles since the fluctuations in temperature and humidity are
larger in the lower levels of the atmosphere (below 4~km) than in the upper
layers. Consequently, a more precise description of actual atmospheric
conditions with GDAS will alter the energy reconstruction compared with
nMMM-based reconstructions for air showers that penetrate deeply into the
atmosphere, usually high-energy events.  The full atmosphere-dependent
fluorescence calculation alters the light yield for conditions with very low
temperatures, corresponding to higher altitudes. The energy dependence of the
\xmax differences is a combined effect of slightly changed humidity conditions
close to ground and temperature conditions higher up in the atmosphere together
with the full atmosphere-dependent fluorescence calculation
(Fig.~\ref{fig:delta}, middle right).

The difference in energy is quite uniform throughout the year, see
Fig.~\ref{fig:delta}, bottom left. For switching on GDAS instead of nMMM (black
dots), it is confirmed that GDAS describes the conditions at the Auger
Observatory very well, as good as the nMMM. Switching on the full
atmosphere-dependent fluorescence calculation does not show a monthly dependence
because the overall integral of the longitudinal light profiles is hardly
changed, see e.g.~\cite{Keilhauer:2008}. Only the modification of the shape of
the longitudinal light profile causes a small monthly dependence of \xmax
(Fig.~\ref{fig:delta}, bottom right).

\begin{figure}[!t]
  \begin{minipage}[t]{.49\textwidth}
    \centering
    \includegraphics*[width=.99\linewidth,clip]{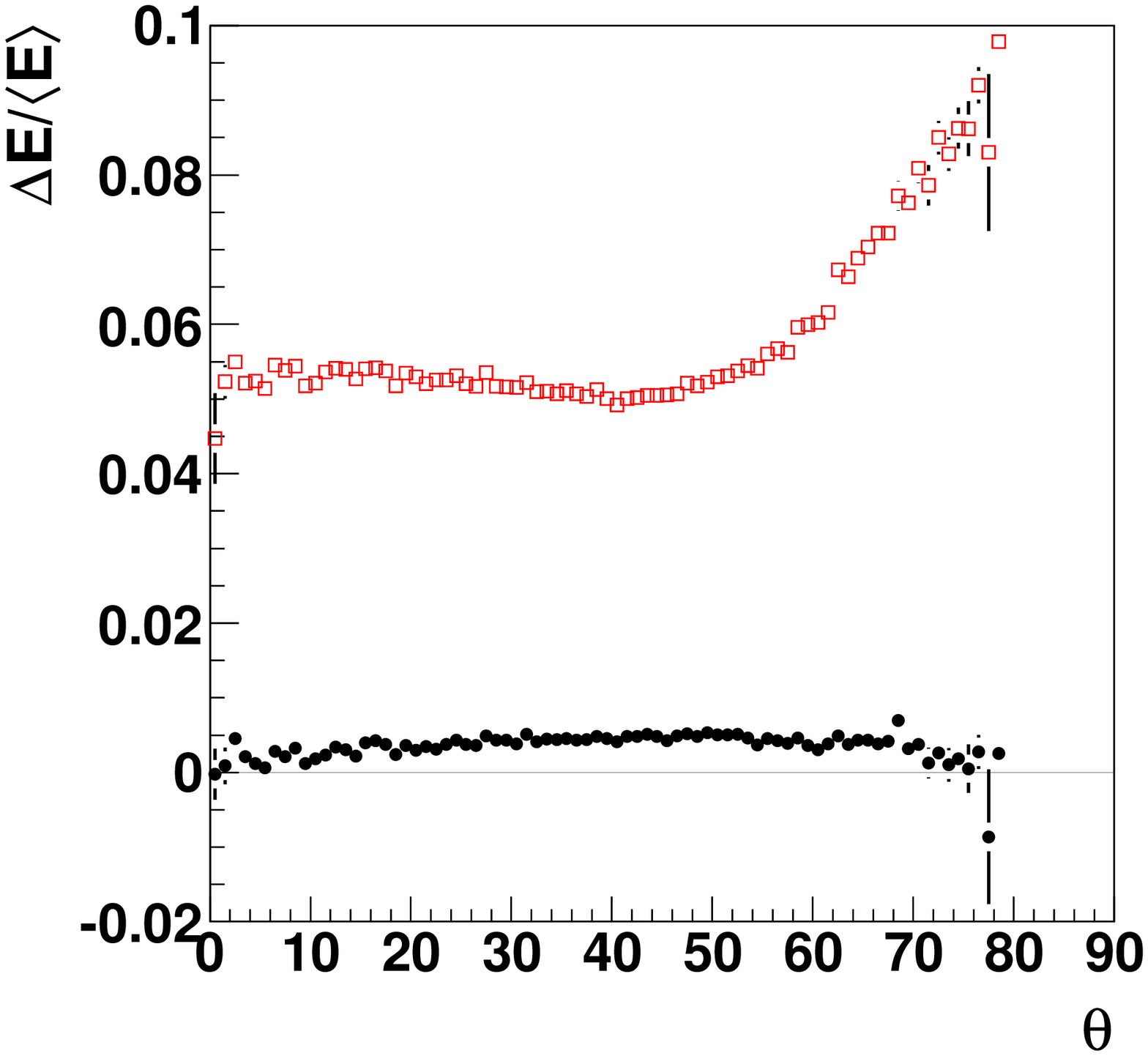}
  \end{minipage}
  \hfill
  \begin{minipage}[t]{.49\textwidth}
    \centering
    \includegraphics*[width=.99\linewidth,clip]{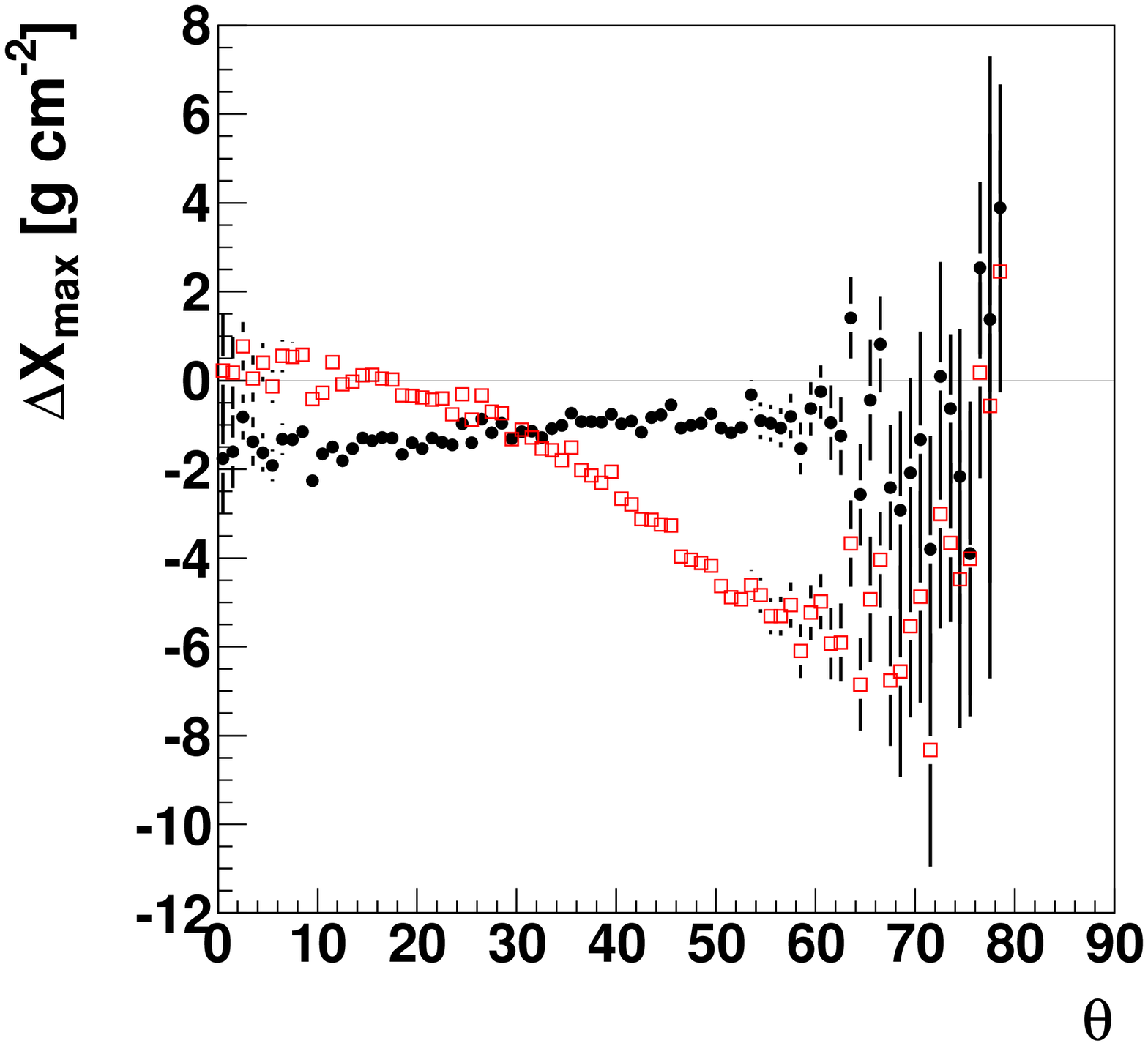}
  \end{minipage}
  \begin{minipage}[t]{.49\textwidth}
    \centering
    \includegraphics*[width=.99\linewidth,clip]{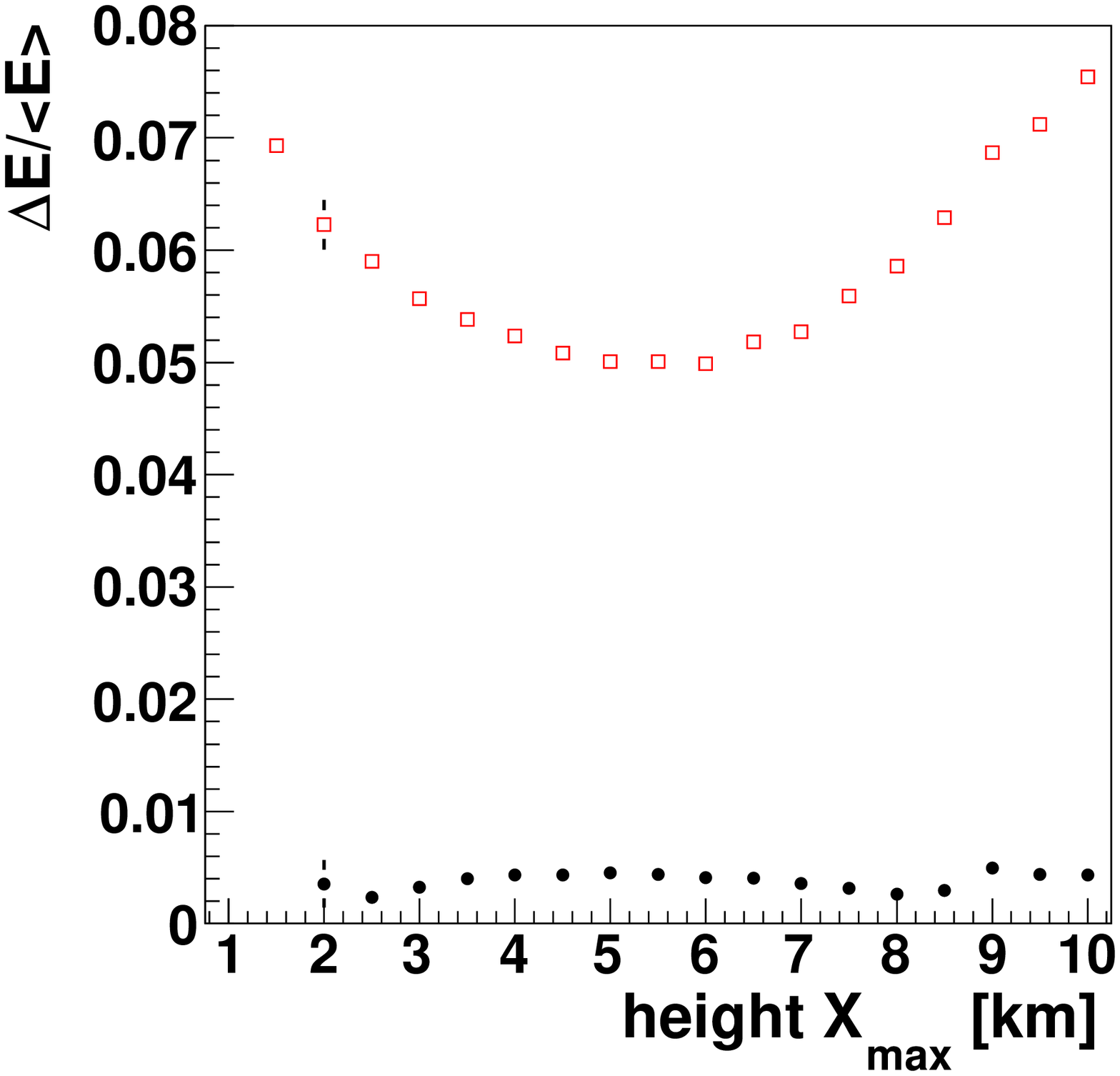}
  \end{minipage}
  \hfill
  \begin{minipage}[t]{.49\textwidth}
    \centering
    \includegraphics*[width=.99\linewidth,clip]{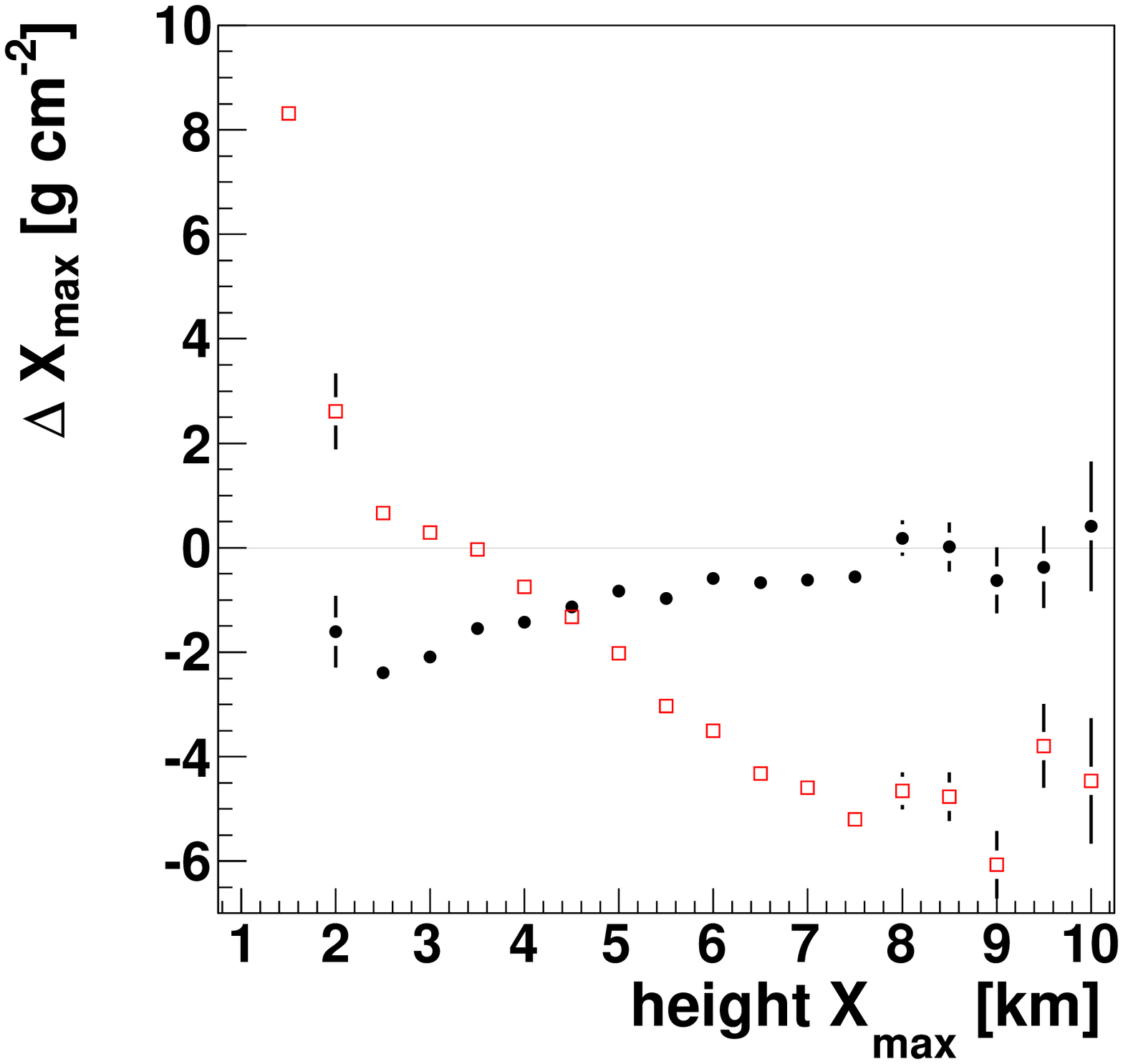}
  \end{minipage}
  \caption{\label{fig:showergeom}
    Energy difference (left) and \xmax difference (right) vs.\ zenith angle
    $\theta$ of EAS (top) and vs.\ geometrical height of shower maximum
    (bottom). Black filled dots are for \textsf{FY$_{\rm mod}^{\rm GDAS}$} minus
    \textsf{FY$_{\rm mod}$}, and red open squares for \textsf{FY$_{\rm mod}^{\rm
    GDAS}$} minus \textsf{FY}.
  }
  \hfill
\end{figure}

In the following, some systematics caused by the particular shower geometry are
studied. In the first set of figures (Fig.~\ref{fig:showergeom}, top row), the
difference in $E$ and \xmax vs.\ zenith angle of the shower $\theta$ is
displayed.  The energy variation is quite uniform around the mean value up to
about 60$^{\circ}$. Only more inclined showers show a stronger shift of
reconstructed energy for the modified fluorescence yield calculation. Concerning
the position of shower maximum, a dependence on zenith angle can be seen above
30$^{\circ}$.

Shown in the bottom row of Fig.~\ref{fig:showergeom}, the $E$ and \xmax
dependence on geometrical height of shower maximum gives a more pronounced view
of the atmospheric conditions in combination with the atmosphere-dependent
fluorescence description. Showers reaching their maximum at an average altitude
between 3 and 7~km show the mean $E$ difference as expected from
Fig.~\ref{fig:delta}, top left. However, showers with very shallow or very deep
positions of shower maximum are reconstructed with a 7--8\% higher primary
energy compared with that using the standard fluorescence calculation. The
reconstructed \xmax follows the expectations according to the study shown in
\cite{Keilhauer:2008}.

\subsection{Impact on Shower Reconstruction Uncertainties}

To study the effect that the GDAS data have on the uncertainties of air shower
reconstructions, air showers induced by protons and iron nuclei are simulated
using the CONEX shower generator~\cite{CONEX} with the QGSJETII hadronic
interaction model~\cite{QGSJETII} for shower energies between 10$^{17.5}$~eV and
10$^{20}$~eV. The fluorescence light is generated including
temperature-dependent collisional cross sections and vapor quenching. The time
stamps of the air shower events correspond to the times of 109~radio soundings
between August 2002 and December 2008 so that actual atmospheric profiles can be
used in the simulation. All 109~launches were performed at night during
cloud-free conditions. After the atmospheric transmission, the detector optics
and electronics are simulated. The resulting data are then reconstructed using
the radiosonde data, as well as the GDAS data.

A basic set of quality cuts is applied. The shower maximum has to be in the
observed part of the track, and the uncertainty in reconstructed energy and
\xmax must be below 20\% and 40~\gcmsq, respectively. Also, the Gaisser-Hillas
profile fit has to have a given quality, $\chi^2$/Ndf~$<$~2, and the fraction of
Cherenkov light from the shower observed by the telescopes must be less than
50\%. After applying all cuts, the values of energy and \xmax from the
reconstructions with different atmospheric description are compared.  The
differences in these reconstructions yield the uncertainties that are introduced
by the use of the GDAS data instead of the actual atmospheric profile.

The same study has been performed to determine the uncertainties from the
nMMM~\cite{Keilhauer:2009icrc2}. There, the systematic error is less than 1\% in
energy and less than 2~\gcmsq in \xmax. In the energy range from 10$^{17.5}$~eV
to 10$^{20}$~eV, random energy-dependent reconstruction uncertainties of
$\pm$1\% and $\pm$5~\gcmsq for low energies up to $\pm$2\% and $\pm$7~\gcmsq for
high energies were found. In the course of our new study, we compute the same
uncertainties due to the nMMM again. The main difference between the current
analysis and the previous one is the implementation of the temperature-dependent
collisional cross sections and the humidity quenching in the calculation of the
fluorescence yield.

\begin{figure}[!t]
  \begin{minipage}[t]{.49\textwidth}
    \centering
    \includegraphics*[width=.99\linewidth,clip]{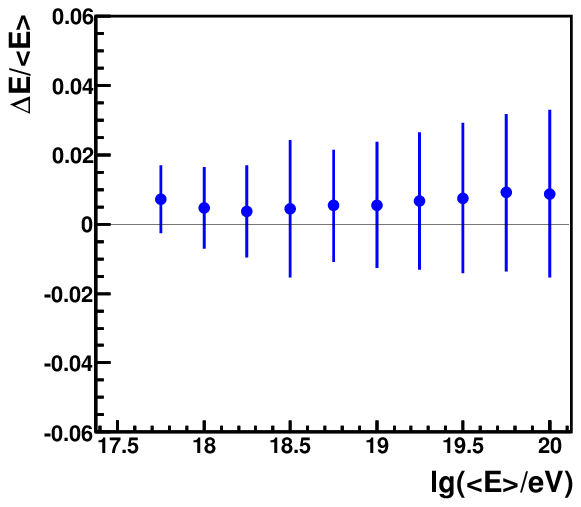}
  \end{minipage}
  \hfill
  \begin{minipage}[t]{.49\textwidth}
    \centering
    \includegraphics*[width=.99\linewidth,clip]{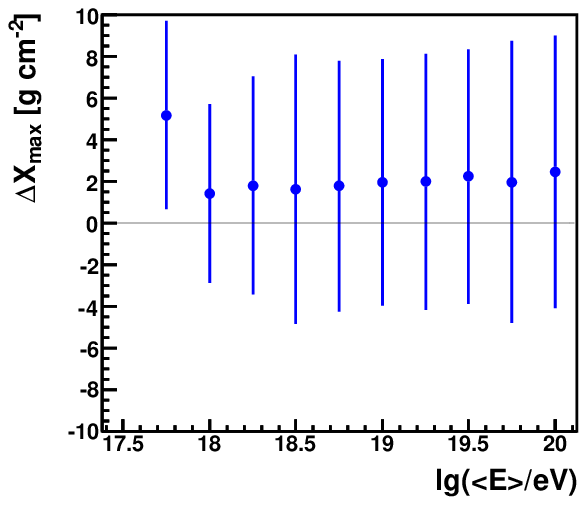}
  \end{minipage}
  \begin{minipage}[t]{.49\textwidth}
    \centering
    \includegraphics*[width=.99\linewidth,clip]{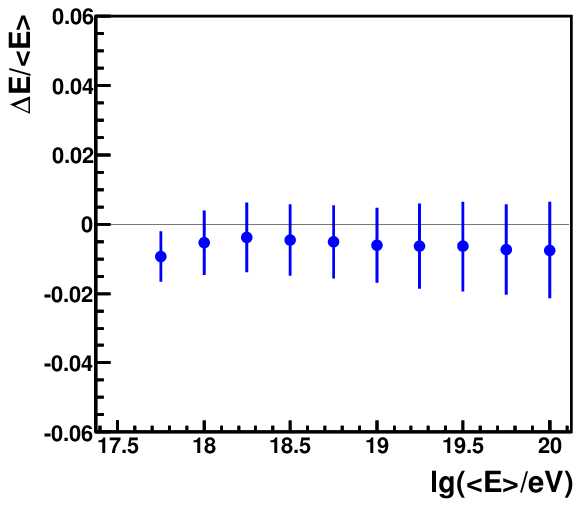}
  \end{minipage}
  \hfill
  \begin{minipage}[t]{.49\textwidth}
    \centering
    \includegraphics*[width=.99\linewidth,clip]{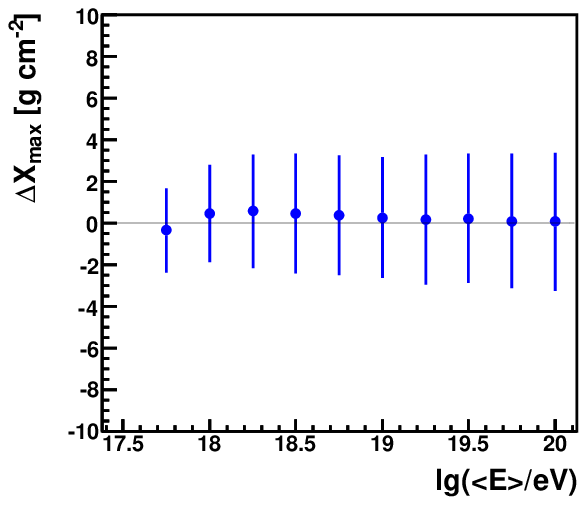}
  \end{minipage}
  \caption{\label{fig:sim_uncert}
    Energy difference (left) and \xmax difference (right) vs.\ reconstructed FD
    energy for simulated showers. The influences of the monthly
    models (top) and of GDAS data (bottom) are shown. Error bars denote
    the RMS spread.
  }
  \hfill
\end{figure}

In Fig.~\ref{fig:sim_uncert}, both results of the new study are shown, the
uncertainties due to nMMM and those due to GDAS. A deviation from zero indicates
a systematic error, and the error bars denote the RMS spread of all simulated
events and are a measure of the reconstruction uncertainty due to this
atmospheric parameterization. In the top panels, the influence of the nMMM
compared to the real atmospheric parameters from radiosondes is shown. The
results for the systematic shifts are in agreement with the previous
study~\cite{Keilhauer:2009icrc2}, with only the systematic shift for \xmax in
the lowest energy bin being higher. The RMS spread in energy is $\pm$0.9\% for
low energies and up to $\pm$2.4\% for high energies. In \xmax, the RMS is
between $\pm$4~\gcmsq for low energies and $\pm$6.5~\gcmsq for high energies.

The influence on the reconstruction due to GDAS data is shown in the bottom part
of Fig.~\ref{fig:sim_uncert}. The systematic shifts in energy are of the same
order, below 1\%, but of opposite sign. The shifts in \xmax are much smaller
than for the nMMM, less than 0.5~\gcmsq. The RMS spread is also considerably
smaller, $\pm$0.9\% and $\pm$2~\gcmsq for low energies, and $\pm$1.3\% and
$\pm$3.5~\gcmsq for high energies. The energy uncertainty at low energies is
comparable to the uncertainty introduced by the nMMM, but at high energies the
uncertainty is reduced by almost 50\%. For \xmax, the uncertainties in all
energy bins are halved.

This study of the reconstruction uncertainties further demonstrates the
advantages of GDAS data over the monthly mean profiles.

\section{Conclusion\label{sec:conclusion}}

The reconstructions of air showers measured at the Pierre Auger Observatory have
used a set of monthly mean profiles as the standard atmospheric description
until recently.  These profiles are averages from meteorological radio soundings
performed at the site of the observatory over several years. The mean profiles
describe the local conditions reasonably well, but cannot describe short-term
variations in the atmosphere. Because of the large burden radio soundings impose
on the collaboration, and their ambiguous duration of validity, data from the
Global Data Assimilation System (GDAS) are a welcome substitute for atmospheric
descriptions. GDAS data rely on established meteorological models and have an
excellent time resolution of 3~hours.

A direct comparison of GDAS data for the site of the Auger Observatory with
local atmospheric measurements validates the adequate accuracy of GDAS data with
respect to horizontal and vertical as well as temporal resolution. The suitable
online publication of these data by NCEP allows for an easy and timely updating
of atmospheric databases used at the Pierre Auger Observatory.

With an air shower reconstruction analysis, the applicability of GDAS data to
air shower analyses can be confirmed along with an improved accuracy with
respect to atmospheric conditions. Also, the value of using an
atmosphere-dependent fluorescence description has been demonstrated. Using
simulated air showers, we show that the GDAS data significantly reduce the
systematic errors and overall uncertainties in air shower reconstructions.

Because of the results discussed in this study, the standard air shower analyses
of the Pierre Auger Observatory are now applying atmospheric data from GDAS and
the fluorescence description \textsf{FY$_{ \rm mod}^{\rm GDAS}$} as described in
Sec.~\ref{sec:reco}.

\section*{Acknowledgments}
The successful installation, commissioning and operation of the Pierre Auger
Observatory would not have been possible without the strong commitment and
effort from the technical and administrative staff in Malarg\"ue.

We are very grateful to the following agencies and organizations for financial support:
Comisi\'on Nacional de Energ\'{\i}a At\'omica,
Fundaci\'on Antorchas,
Gobierno De La Provincia de Mendoza,
Municipalidad de Malarg\"ue,
NDM Holdings and Valle Las Le\~nas, in gratitude for their continuing
cooperation over land access, Argentina;
the Australian Research Council;
Conselho Nacional de Desenvolvimento Cient\'{\i}fico e Tecnol\'ogico (CNPq),
Financiadora de Estudos e Projetos (FINEP),
Funda\c{c}\~ao de Amparo \`a Pesquisa do Estado de Rio de Janeiro (FAPERJ),
Funda\c{c}\~ao de Amparo \`a Pesquisa do Estado de S\~ao Paulo (FAPESP),
Minist\'erio de Ci\^{e}ncia e Tecnologia (MCT), Brazil;
AVCR, AV0Z10100502 and AV0Z10100522,
GAAV KJB300100801 and KJB100100904,
MSMT-CR LA08016, LC527, 1M06002, and MSM0021620859, Czech Republic;
Centre de Calcul IN2P3/CNRS,
Centre National de la Recherche Scientifique (CNRS),
Conseil R\'egional Ile-de-France,
D\'epartement  Physique Nucl\'eaire et Corpusculaire (PNC-IN2P3/CNRS),
D\'epartement Sciences de l'Univers (SDU-INSU/CNRS), France;
Bun\-des\-mi\-ni\-ste\-ri\-um f\"ur Bil\-dung und For\-schung (BMBF),
Deut\-sche For\-schungs\-ge\-mein\-schaft (DFG),
Fi\-nanz\-mi\-ni\-ste\-ri\-um Ba\-den-W\"urt\-tem\-berg,
Helm\-holtz-Ge\-mein\-schaft Deut\-scher For\-schungs\-zen\-tren (HGF),
Mi\-ni\-ste\-ri\-um f\"ur Wis\-sen\-schaft und For\-schung, Nord\-rhein-West\-falen,
Mi\-ni\-ste\-ri\-um f\"ur Wis\-sen\-schaft, For\-schung und Kunst, Ba\-den-W\"urt\-tem\-berg, Germany;
Istituto Nazionale di Fisica Nucleare (INFN),
Istituto Nazionale di Astrofisica (INAF),
Ministero dell'Istruzione, dell'Universit\`a e della Ricerca (MIUR),
Gran Sasso Center for Astroparticle Physics (CFA), Italy;
Consejo Nacional de Ciencia y Tecnolog\'{\i}a (CONACYT), Mexico;
Ministerie van Onderwijs, Cultuur en Wetenschap,
Nederlandse Organisatie voor Wetenschappelijk Onderzoek (NWO),
Stichting voor Fundamenteel Onderzoek der Materie (FOM), Netherlands;
Ministry of Science and Higher Education,
Grant Nos. 1 P03 D 014 30 and N N202 207238, Poland;
Funda\c{c}\~ao para a Ci\^{e}ncia e a Tecnologia, Portugal;
Ministry for Higher Education, Science, and Technology,
Slovenian Research Agency, Slovenia;
Comunidad de Madrid,
Consejer\'{\i}a de Educaci\'on de la Comunidad de Castilla La Mancha,
FEDER funds,
Ministerio de Ciencia e Innovaci\'on and Consolider-Ingenio 2010 (CPAN),
Generalitat Valenciana,
Junta de Andaluc\'{\i}a,
Xunta de Galicia, Spain;
Science and Technology Facilities Council, United Kingdom;
Department of Energy, Contract Nos. DE-AC02-07CH11359, DE-FR02-04ER41300,
National Science Foundation, Grant No. 0969400,
The Grainger Foundation USA;
ALFA-EC / HELEN,
European Union 6th Framework Program,
Grant No. MEIF-CT-2005-025057,
European Union 7th Framework Program, Grant No. PIEF-GA-2008-220240,
and UNESCO.


\bibliography{gdas_paper}
\bibliographystyle{elsarticle-num}

\clearpage

\end{document}